\documentclass[prb,twocolumn,floatfix,superscriptaddress]{revtex4}
\usepackage{graphicx}
\usepackage{amssymb,bm}

\begin{document}
\title{Impurity and correlation effects on transport in 
one-dimensional quantum wires}
\author{T.\ Enss}
\affiliation{Max-Planck-Institut f\"ur Festk\"orperforschung,
  Heisenbergstr.\ 1, D-70569 Stuttgart, Germany}
\author{V.\ Meden}
\affiliation{Institut f\"ur Theoretische Physik, Universit\"at G\"ottingen, 
Friedrich-Hund-Platz 1, D-37077 G\"ottingen, Germany}
\author{S.\ Andergassen}
\affiliation{Max-Planck-Institut f\"ur Festk\"orperforschung,
  Heisenbergstr.\ 1, D-70569 Stuttgart, Germany}
\author{X.\ Barnab\'e-Th\'eriault}\thanks{Xavier B.-T. passed away in a tragic 
traffic accident on August 15, 2004.}
\affiliation{Institut f\"ur Theoretische Physik, Universit\"at G\"ottingen, 
Friedrich-Hund-Platz 1, D-37077 G\"ottingen, Germany}
\author{W.\ Metzner}
\affiliation{Max-Planck-Institut f\"ur Festk\"orperforschung,
  Heisenbergstr.\ 1, D-70569 Stuttgart, Germany}
\author{K.\ Sch\"onhammer}
\affiliation{Institut f\"ur Theoretische Physik, Universit\"at G\"ottingen, 
Friedrich-Hund-Platz 1, D-37077 G\"ottingen, Germany}

\begin{abstract}
We study transport through a one-dimensional quantum wire  
of correlated fermions connected to semi-infinite leads. 
The wire contains either a single impurity or two  
barriers, the latter allowing for resonant tunneling. 
In the leads the fermions are 
assumed to be non-interacting. The wire is described by a microscopic 
lattice model. Using the functional renormalization group we 
calculate the linear conductance for wires of mesoscopic 
length and for all relevant temperature scales.
For a single impurity, either strong or weak, we find  power-law 
behavior as a function of temperature. In addition, we can describe 
the complete crossover from the 
weak- to the strong-impurity limit. For two barriers, depending on 
the parameters of the enclosed quantum dot, we find temperature
regimes in which the conductance 
follows power-laws with ``universal'' exponents as well as non-universal 
behavior. Our approach leads to a comprehensive picture of resonant 
tunneling. We compare our results with those of alternative approaches.  
\end{abstract}
\maketitle     

\section{Introduction}
\label{Intro}

The interplay of static impurities and correlations in one-dimensional (1d)
Fermi systems leads to a variety of surprising effects. A detailed 
understanding of the physics involved is interesting from the
basic-science point of view as well as from the perspective 
of a possible application of 1d quantum wires in future nano-electronics. 
In a 1d metallic system, correlations have a strong effect on the 
low-energy properties. Quite differently from the conventional Fermi-liquid
behavior and even in the homogeneous case, physical properties at low 
energy scales follow power-laws, described by the Tomonaga-Luttinger 
liquid (TLL) phenomenology.\cite{KS} For spin-rotational invariant and
spinless models the exponents can be expressed in terms of a single
number, the interaction-dependent TLL 
parameter $K$.  

A single static impurity in a TLL with repulsive interaction 
($0<K<1$) leads to a dramatic modification of  the low-energy physics
as is obvious from lowest order perturbation theory in the
strength of the impurity.\cite{LutherPeschel,Mattis,ApelRice,Giamarchi} 
More elaborate methods show that for vanishing energy
scale $s$ the conductance of such a system vanishes following a 
power-law.\cite{KaneFisher,Furusaki0,Moon,MatveevGlazman,Fendley} The scaling 
exponent $2 \alpha_B$ is independent of 
the bare impurity strength and determined by the exponent $\alpha_B =1/K-1$
(for spinless fermions) of the
one-particle spectral weight of a TLL close to an open
boundary.\cite{KaneFisher} 
For weak impurities the correction to the impurity free conductance
scales as $s^{2(K-1)}$, which holds as long as the correction stays
small. Assuming an infinite system the asymptotic
low-energy properties have been investigated intensively within 
an integrable field theoretical model --- the local sine-Gordon 
model (LSGM).\cite{KaneFisher,Furusaki0,Moon,Fendley} For this model,
with fixed $K$ the conductance as a function of temperature follows   
one-parameter scaling for different strengths of the impurity. 
Similar results were obtained for
a TLL connected by arbitrarily ``smooth'' (spatially adiabatic) 
contacts to semi-infinite Fermi-liquid
leads.\cite{Safi,MaslovStone,Ponomarenko,Maslov,FuNa,Egger}  

Also transport through a double barrier enclosing a
quantum dot has been studied over 
the last few years using field theoretical 
models.\cite{KaneFisher,Furusaki1,Chamon,Safi2,Furusaki2,Thorwart,Kleinmann,Komnik,Nazarov,Polyakov,Huegle,Thorwart2}
Tuning the dot energies by a gate voltage $V_g$ resonant
tunneling can be achieved. For appropriately chosen dot parameters 
(dot size, barrier height) and as a function of temperature $T$ the
conductance $G$ shows different temperature regimes with
distinctive power-law scaling and exponents which can be expressed in
terms of
$K$.\cite{Furusaki1,Safi2,Furusaki2,Thorwart,Nazarov,Polyakov,Huegle,Thorwart2}     
For the double-barrier problem no exactly solvable generic model is
known and applying different approximate 
analytical\cite{Furusaki1,Safi2,Furusaki2,Thorwart,Nazarov,Polyakov,Thorwart2} and 
numerical\cite{Huegle} methods has not provided a consistent
picture. 

To study the effect of the interplay of correlations and impurities 
on transport 
on all relevant energy scales 
we use the functional renormalization group (fRG) 
method. 
The fRG can be applied directly to microscopic models.
We consider the lattice 
model of spinless fermions with nearest-neighbor hopping $t$ 
and nearest-neighbor interaction $U$. The impurities are 
modeled by either locally raising site energies or reducing hopping 
matrix elements  across bonds. The fRG was recently 
introduced\cite{Manfred} as a new powerful tool for studying
interacting Fermi systems. It provides a systematic
way of resumming competing instabilities\cite{2dsystems} and goes
beyond simple perturbation theory even in problems which are not 
plagued by infrared divergences.\cite{Ralf}
The fRG procedure 
we use starts from an exact hierarchy of differential flow
equations for the one-particle irreducible vertex
functions,\cite{Wetterich,Morris,SalmhoferHonerkamp} 
as e.g.\ the self-energy and the effective two-particle interaction.  
It is derived by replacing the free propagator by a propagator depending on
an infrared cutoff $\Lambda$ and taking the derivative of the generating 
functional with respect to $\Lambda$. 

At $T=0$ we introduced earlier two truncation 
schemes which led to a manageable number of coupled 
equations.\cite{VM1,VM2,Sabine} The flow of the self-energy, 
which in particular encodes the renormalized impurity potential, 
is fully taken into account, while 
the two-particle vertex is parametrized by a  
single flowing coupling constant.\cite{Sabine}  The 
bare two-particle 
interaction $U$ is taken as a small parameter, but the impurities 
can have arbitrary strength and shape. We are 
thus in a position to study the single- and double-barrier
problem by applying the same approximate method. 
For weak and strong single impurities the flow equations can be solved
analytically.\cite{VM2} The results are consistent with the above
mentioned power-law scaling of the conductance. For general
parameters, in particular in the double barrier case, we 
solved the coupled differential equations numerically which at 
$T=0$ can easily be done for systems of up to $N=10^7$ lattice 
sites. Using the fRG we previously studied the
one-particle spectral function close to a single 
impurity,\cite{VM1,VM2,Sabine} the decay of density (Friedel-)
oscillations off an impurity,\cite{Sabine} and the persistent current 
in mesoscopic rings pierced by a magnetic flux.\cite{VM3} For these 
observables we recovered the expected ``universal'' power-laws with
exponents which for $1/2 \leq K < 1$  ($K \to 1$ for $U \to 0$) 
turned out to be in good agreement with the ones known from the
LSGM. In addition it was shown that the asymptotic behavior typically
sets in on very small energy scales and for very large systems, except
for very strong bare impurities. Within our approximation, in the 
calculation of various observables we have to solve as a last step the 
problem of a non-interacting fermion moving in the renormalized impurity 
potential. This mapping on an effective one-particle problem is helpful
for the understanding of the observed effects.

In a short publication we previously discussed the transport for the
single-barrier case.\cite{VM4} The results were obtained after 
supplementing the $N$-site wire of interacting fermions (called
the ``interacting wire'' in the following) by  
1d semi-infinite leads.
We studied the $T=0$ behavior as a function of $1/N$.
For ``smooth'' contacts between the leads and 
the interacting wire our data for different impurity strengths
and inverse system sizes collapse onto a single curve 
after applying a one-parameter scaling ansatz. 
This shows that even a
small bare impurity leads to a vanishing conductance in the limit $N
\to \infty$.
At $K=1/2$ the one-parameter scaling function of the infinite LSGM 
is known analytically.\cite{KaneFisher,Fendley} 
Up to a small error our rescaled data fall onto this curve. 
As an additional benchmark  
for small $N$ we compared the fRG approximation for the conductance 
with results of density-matrix renormalization group (DMRG)
calculations.\cite{VM4,VM5} These tests lead us to conclude that our 
approximate method works reliably for interactions in the range
$1/2 \leq  K \leq 1$. In addition, the collapse of our 
fRG data onto the $K=1/2$ local 
sine-Gordon scaling function provides an indication that the low-energy 
physics of the LSGM can indeed be found 
in a large class of models of inhomogeneous, one-dimensional, and 
correlated electron systems, which is not obvious {\it a priori.}

In the present publication we extend the fRG scheme of 
Ref.\ \onlinecite{Sabine} to finite temperatures.
As in any experimental 
system the bandwidth $B$ and the inverse of the wire length set upper 
and lower bounds for the possible  ``universal'' scaling behavior
of the conductance.
This has to be contrasted to studies of field-theoretical models 
with $B =\infty$ and (in most cases) $1/N=0$. We calculate $G$ 
on all temperature scales.
Depending on the bare impurity and interaction strength an asymptotic
low-energy regime might not be reachable in experiments on finite wires.
For finite $T$ we consider up to $N=10^4$ sites. For
typical lattice constants this leads to interacting wires in the 
micrometer range, roughly corresponding to  
quantum wires accessible
to transport experiments. Within the last few years, transport
properties were  
measured in different, effectively 1d mesoscopic 
wires.\cite{Tarucha,Bockrath,Yao,Auslaender,Fuhrer2,Papadopoulous,Picciotto,Postma}

Although in many respects our model is more realistic 
than the field-theoretical models studied so far, important 
ingredients for an appropriate description of the experimental 
situation are missing, as e.g.\ the spin degree 
of freedom, higher-dimensional leads, and realistic contacts. 
In the present paper we therefore refrain from a detailed comparison of 
our results to experiments. Later, we  briefly 
comment on certain aspects of the experimental results and 
discuss the possibility of extending our method to treat 
models closer to  experimental systems. 

In the two limits
of weak and strong impurities we recover the expected power-law
scaling of $G(T)$ with exponents which are in good agreement with the ones
known from the LSGM. 
We show that one-parameter 
scaling is achieved. 
We also present results for attractive interactions
with $U<0$ and $K > 1$.
Based on our findings we discuss the accuracy of our approximation 
scheme in detail. We investigate the relation between our 
approach and a ``leading-log'' resummation for the effective 
transmission at the chemical potential introduced in 
Ref.\ \onlinecite{MatveevGlazman}.
We then consider 
the double-barrier problem. 
Within a single approximation scheme all relevant temperature
regimes can be investigated, leading to a 
comprehensive picture. 
We compare our findings to results 
that have been obtained for restricted temperature regimes by 
applying alternative 
approximate techniques and a numerical method to field-theoretical models. 
A brief account of results for resonant 
tunneling in the double-barrier problem obtained from the fRG 
has been presented earlier, focusing on the scaling of the peak 
conductance for symmetric barriers.\cite{VM6}

This paper is organized as follows. In Sect.\
\ref{model} we introduce the model considered. 
In Sect.\ \ref{Scattheory} results 
of single-particle scattering theory are derived. They are useful for the 
understanding of our findings for the conductance.
The details of the fRG
method are presented in Sect.\ \ref{fRGintransport}. We  discuss
the general approximation scheme and describe how the fRG is extended to
finite $T$ and to the study of transport.  In Sect.\ 
\ref{Results} we present our results for the transport in the presence 
of a single impurity as well as through a double barrier. We conclude  
with a summary and an outlook in Sect.\ \ref{Summary}. In the appendix 
we briefly describe the relation between the imaginary part of the
self-energy and the occurrence of current-vertex corrections in a 
Landauer-B\"uttiker-type expression for the conductance. 
 
\section{The model}
\label{model}

We consider spinless fermions on a lattice
with nearest-neighbor interaction supplemented by various types 
of impurities. The Hamiltonian reads
\begin{equation}
\label{fullH}
 H = H_{\rm kin} + H_{\rm int} + H_{\rm imp} \; .
\end{equation}
The kinetic energy is given by nearest-neighbor hopping with an
amplitude $t$,
\begin{equation}
\label{kindef}
 H_{\rm kin} = -t \sum_{j=-\infty}^{\infty} \big( \,
 c^{\dag}_{j+1} c_j^{\phantom\dag} + c^{\dag}_j \, c_{j+1}^{\phantom\dag}
 \, \big) \; , 
\end{equation}
where we used standard second-quantized notation with $c^{\dag}_j$ 
and $c_j^{\phantom\dag}$ being creation and annihilation operators on site $j$,
respectively. 

The part of the Hamiltonian containing the interaction reads 
\begin{eqnarray}
\label{intdef}
H_{\rm int} & = & \sum_{j=1}^{N-1} U_{j,j+1}  \left[ n_j - \nu(n,U)
\right]  \left[ n_{j+1} - \nu(n,U) \right] ,
\end{eqnarray}
with the local density operator $n_j = c^{\dag}_j \,
c_j^{\phantom\dag}$. The interaction acts
only between the bonds of the sites $1$ to $N$, which define the
interacting wire. The interaction $U_{j,j+1}$ between electrons on sites $j$
and $j+1$ is allowed to depend on the position. 
 As discussed in Refs.\ \onlinecite{VM4} and 
\onlinecite{VM5}  in the absence of a one-particle impurity potential, 
turning on the interaction sharply leads to a reduction of the conductance 
compared to $e^2/h$.\cite{footnoteleads} As in the other theoretical
studies mentioned we here avoid this effect by modeling ``perfect''
contacts. To achieve this $U_{j,j+1}$ is taken as a smoothly 
increasing function of $j$ starting form zero at the bond $(1,2)$ and 
approaching a constant bulk value $U$ over a sufficiently large number 
of bonds. Equally, the $U_{j,j+1}$ are switched off 
close to the bond $(N-1,N)$.
The results are independent of the detailed shape of the envelope
function as long as it is sufficiently smooth.
The larger  $N$ the ``smoother'' $U_{j,j+1}$ has to be chosen. 
More explicitly we here take ($N$ even, $j=1,\ldots,N/2$)
\begin{eqnarray}
\label{shape}
  U_{j,j+1} = U \frac{\arctan{[s(j-j_s)]}-\arctan{[s(1-j_s)]}}
  {\arctan{[s(N/2-j_s)]}-\arctan{[s(1-j_s)]}} 
\end{eqnarray} 
for the left part of the interacting wire and likewise for the
right. For odd $N$ a similar function is used. 
The parameters $s$ and $j_s$ are chosen such that in the absence of any 
impurity for $T=0$ we find $1-G/(e^2/h) < 10^{-6}$. 
To achieve this for the $N$ considered here 
$j_s = 56$ and $s \approx 1/4$ is sufficient.   
The two regions of the lattice with $j < 1$ and $j>N$
constitute the two semi-infinite leads.

In Eq.~(\ref{intdef}) we shifted $n_j$ by 
a parameter $\nu(n,U)$, which depends on the filling factor $n$ and $U$. 
This is equivalent to introducing an additional  one-particle
potential in the interacting part of the system.
The parameter $\nu(n,U)$ is chosen in such a way that 
the density in the interacting wire acquires the
desired value $n$.

\begin{figure}[htb]
\begin{center}
\includegraphics[width=0.40\textwidth,clip]{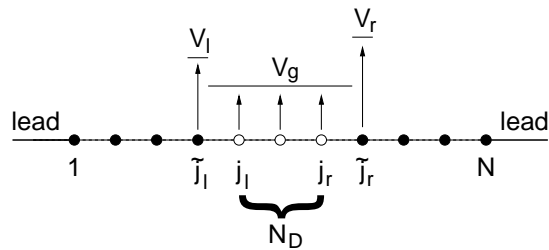}
\end{center}
\vspace{-0.3cm}
\caption[]{Schematic plot of the quantum dot situation, where the
  barriers are modeled by two site impurities. \label{doublebar}}
\end{figure}

The general form of the impurity part of the Hamiltonian is written as
\begin{equation}
\label{impdef}
 H_{\rm imp} = \sum_{j,j'} 
 V_{j,j'}^{\phantom\dag} \; c^{\dag}_{j} \, c_{j'}^{\phantom\dag} \; ,
\end{equation}
where $V_{j,j'}$ is a static potential.
Site impurities are given by a local potential 
\begin{equation}
 V_{j,j'} = V_j \, \delta_{j,j'} \; .
\end{equation}
In our study of the effect of a single barrier we mainly consider a
site impurity located far away from both leads on site 
$j_0$ with $j_0\gg 1$ and $N-j_0\gg 1$,
\begin{equation}
 V_j = V \, \delta_{j,j_0} \; .
\end{equation}
In the discussion of resonant tunneling two site impurities of
strengths $V_l$ and $V_r$ on the sites
$\tilde j_l=j_l-1$ and $\tilde j_r=j_r+1$  
constitute the barriers, with $j_l \gg 1 $
and $N-j_r \gg 1$. The $N_D$ sites between $j_l$ and $j_r$ define
the quantum dot with $j_r - j_l +1 = N_D \geq 1$. 
The effect of a
gate voltage is described by a constant $V_g$ on sites $j_l$ to
$j_r$. 
This situation is depicted in Fig.~\ref{doublebar}.
We also consider hopping impurities  
described by non-local potentials 
\begin{equation}
 V_{j,j'} = V_{j',j} = - t_{j,j+1} \, \delta_{j',j+1} 
\end{equation}
as barriers. For the special case of a single hopping impurity,
\begin{equation}
\label{hoimp}
 t_{j,j+1} = (t'-t) \, \delta_{j,j_0} \; ,
\end{equation}
the hopping amplitude $t$ is replaced by $t'$ on the bond linking
the sites $j_0$ and $j_0+1$. In the double-barrier problem 
we consider a hopping $t_l$ across the bond $(\tilde j_l,j_l)$  and $t_r$
across $(j_r,\tilde j_r)$. 
For small $t_l$ and $t_r$ also the
interactions across the bonds $(\tilde j_l,j_l)$ and $(j_r,\tilde j_r)$,
i.e.\ $U_{\tilde j_{l},j_l}$ and $U_{j_r,\tilde j_r}$, 
are assumed to be reduced. 
In the following we set the bulk hopping amplitude $t$ equal
to one, i.e.\ all energies are given in units of $t$. This leads to
the non-interacting dispersion $\varepsilon_k=-2\cos k$ and 
bandwidth $B=4$. The lattice constant is set to $1$.  

The homogeneous model $H=H_{\rm kin} + H_{\rm int}$ with a constant
interaction $U$ across all bonds (not only the ones within $[1,N]$)
can be solved exactly by Bethe ansatz.\cite{Haldane} 
It shows TLL behavior for all particle densities $n$ 
and any interaction strength except at half filling 
for $|U| > 2$. The $U$- and $n$-dependent 
TLL parameter $K$ can be determined solving coupled
integral equations,\cite{Haldane} which in the half-filled case can be done
analytically with the result 
\begin{equation}
\label{BetheAnsatz}
 K = \left[\frac{2}{\pi} \, 
 \arccos \left(-\frac{U}{2} \right) \right]^{-1} \; ,
\end{equation}
for $|U| \leq 2$.

We compute the conductance of the above models in the linear-response
regime. For $T=0$ the fRG flow equations can
be solved numerically for interacting wires as large as $N=10^7$ lattice
sites. For finite $T$ we treat systems with up to $N=10^4$ 
sites. 

\section{Scattering on one-dimensional lattices}
\label{Scattheory}

The truncation of our fRG flow equations leads to a scattering 
problem of non-interacting fermions in 
an effective potential given by the non-trivial spatial
dependence of the self-energy on sites $1$ to $N$.\cite{VM1,VM2,Sabine}
It is thus useful to present some results for the scattering in a 1d 
system with an arbitrary (non-local) potential in a finite segment of the 
infinite wire. The results are important for the
understanding of how the fRG based approximation scheme is set up and
how the conductance is calculated within this scheme 
(see Sect.~\ref{fRGintransport}). 
In the first subsection we focus on this aspect. In the second
more technical subsection we derive relations which can be used to
gain a deeper understanding of our results for the conductance presented 
in Sect.~\ref{Results}. In order to make 
this section self-contained we include short derivations of well-known 
results, as e.g.\ Eq.\ (\ref{transmission}). 

\subsection{General relations}

The single-particle Hamiltonian $H_{1p}=H_0+V_{\rm LR}$ we consider 
in this section reads
\begin{eqnarray}
H_0 &=& -\sum_{j=-\infty}^0\left (|j-1\rangle \langle j|+ \mbox{h.c.}\right)
       \nonumber \\  && -\sum_{j=N+1}^\infty 
\left( |j\rangle \langle j+1|+ \mbox{h.c.} \right)
       +H_s  \nonumber \\
V_{\rm LR}&=& - t_L\left ( |0\rangle \langle 1| +\mbox{h.c.}\right )
         - t_R\left ( |N\rangle \langle N+1|+\mbox{h.c.}\right )
         \nonumber \, ,
\end{eqnarray}
\vspace{-.6cm}
\begin{equation}
\label{Hamiltonian}
\vspace{-.3cm}
\end{equation}
where $|j \rangle $ denotes the state in which the fermion is 
centered at lattice site $j$. 
Here $H_s$ is a general single-particle 
Hamiltonian in the scattering segment from
site $1$ to $N$ which is connected to ideal leads with
nearest-neighbor hopping $t=1$.
The hopping matrix elements $t_L$ and $t_R$ connecting the three parts
described by $H_0$ are assumed to be real.  

There are two ways to describe the scattering problem. In
the usual approach the ideal infinite system is used as the
unperturbed system and the deviations are treated as the perturbation.
In Eq.\ (\ref{Hamiltonian}) the grouping has been made differently.
The disconnected system ``left lead -- scattering region -- right lead''
is described by $H_0$ and the connections proportional to $t_L$ and $t_R$
present the perturbation in the construction of the scattering
states $ |k,a\pm \rangle $\cite{Taylor}
\begin{eqnarray}
|k,a\pm \rangle &=& \lim_{\eta \to 0} \frac{\pm
  i\eta}{\varepsilon_k-H_{1p} \pm i\eta}
                   |k,a\rangle \nonumber \\
& = &     \lim_{\eta \to 0}\pm i\eta  \;  G(\varepsilon_k\pm i\eta) |k,a\rangle
      \nonumber \\
             &=&    |k,a\rangle+ G(\varepsilon_k\pm i0)V_{\rm LR} |k,a\rangle
             \, . 
\end{eqnarray}
Here the index $a$ stands for $L,R$ and the
unperturbed states $|k,L(R)\rangle $ are
standing waves in the left (right) semi-infinite lead, e.g.\ for $j
\leq 0$
\begin{eqnarray}
\label{standing}
\langle j |k,L\rangle & = & \sqrt{\frac{2}{\pi}}\sin{\left[k(j-1)
  \right]}  \nonumber \\
& = & \frac{-i}{\sqrt{2\pi}}\left[ e^{ik(j-1)}- e^{-ik(j-1)} \right] \; .
\end{eqnarray}
In this section $G(z)$ denotes the resolvent matrix with respect to
$H_{1p}$. With the above normalization of the standing waves, 
$\langle k,a|k',a'\rangle =\delta_{a,a'}
\delta(k-k')$ holds. For $j>N$ the scattering state $|k,L+\rangle $
has components
\begin{eqnarray}
\label{rechts}
\langle j|k,L+\rangle &=& - \langle j| G(\varepsilon_k+i0)|1\rangle
t_L\langle 0|k,L\rangle  \nonumber \\
 &=& \langle j| G_0(\varepsilon_k+i0)|N+1 \rangle t_R \nonumber \\
&& \times  \langle N| G(\varepsilon_k+i0)|1\rangle t_L\langle 0|k,L\rangle.
\end{eqnarray}
In the second equation we used $G=G_0+ G_0 V_{\rm LR}G  $ with
$G_0(z)=(z-H_0)^{-1}$.
The resolvent matrix element of the semi-infinite chain in
Eq.\ (\ref{rechts}) is given by\cite{residue} 
\begin{eqnarray*}
  \langle j| G_0(\varepsilon_k+i0)|N+1 \rangle=-e^{ik(j-N)} \; ,
\end{eqnarray*}
 which shows that $\langle j|k,L+\rangle $ is in fact an outgoing
plane wave. From Eqs.\ (\ref{standing}) and (\ref{rechts}) the
transmission amplitude $t(\varepsilon_k)$ can be read off. The
corresponding probability is given by
\begin{equation} 
\label{transmission}
|t(\varepsilon_k)|^2=4t_L^2t_R^2\sin^2k \; |\langle N|
G(\varepsilon_k+i0)|1\rangle|^2 \; .
\end{equation}
This is a very important relation,
as in the Landauer-B\"uttiker formula\cite{LanBue}
the transmission probability directly determines the linear conductance
$G(T)$ of non-interacting fermions at temperature $T$. For spinless fermions 
on the lattice it reads 
\begin{equation} 
\label{LB}
G(T)=\frac{e^2}{h}\int_{-B/2}^{B/2}
\left(-\frac{df}{d \varepsilon}\right)|t(\varepsilon)|^2 d\varepsilon,
\end{equation}
where $f(\varepsilon)=1/(e^{\beta (\varepsilon-\mu) }+1)$
is the Fermi function
with $\beta=1/T$ and $\mu$ the chemical potential. The Boltzmann
constant $k_B$ is set to 1.
If the fRG flow is integrated for a finite temperature and a finite
interacting wire 
the self-energy is $T$- and $N$-dependent, i.e.\ $H_s$ and therefore also
the transmission probability $|t(\varepsilon)|^2$ depends on
temperature and wire length. In the notation used in this section these 
additional dependences are left implicit.

For the further discussion of the resolvent matrix element
in Eq.\ (\ref{transmission})
the relation\cite{Taylor}
\begin{eqnarray}
\label{Loewdin}
PG(z)P  & = &  \\ 
&& \hspace{-2.0cm}\left [ zP-PH_{1p}P-PH_{1p}Q\left 
(zQ-QH_{1p}Q \right )^{-1}QH_{1p}P\right ]^{-1} \nonumber \, ,
\end{eqnarray}
where $P$ and $Q$ are projectors with $P+Q=\mathbf 1$, is used
several times.
With this projection formula 
the calculation of $ \langle N| G(\varepsilon_k+i0)|1\rangle $ for the
infinite system can be reduced to the problem of inverting an
$N\times N$-matrix. With the definition $P_s= \sum_{j=1}^N
|j\rangle \langle j|$ one obtains
\begin{eqnarray}
\label{projection}
 \langle N| G(z)|1\rangle =  
\langle N|\left [zP_s-H_s^{\rm eff}(z)\right ]^{-1}  |1\rangle    ,
\end{eqnarray}
where $G_b^0$ is the diagonal element of the resolvent of the
semi-infinite chain at the boundary and 
\begin{eqnarray*}
H_s^{\rm eff}(z) = H_s+G_b^0(z) \left (
t_L^2 |1\rangle\langle 1|+t_R^2|N\rangle\langle N|\right ) \; .
\end{eqnarray*} 
For energies inside the band $G_b^0$ is given by   
\begin{eqnarray*}
G_b^0(\varepsilon+i0)=-e^{ik(\varepsilon)}=
(\varepsilon-i\sqrt{4-\varepsilon^2})/2 \; .
\end{eqnarray*}

In our fRG
description of interacting fermions the ``scattering
segment''  $[1,N]$ is defined by the part of the wire where interactions
are present. 

\subsection{The double barrier problem}

As one of the problems in our paper we discuss transport through a 
quantum dot with $N_D$ sites extending from
$j_l \gg 1$ to $j_r$ with $N-j_r \gg 1$, where $j_r=j_l+N_D-1$.
For the sites neighboring the dot we use the indices $\tilde j_a$ 
($a=l,r$). 
In $H_s$ the (real) hopping matrix elements from site $\tilde j_l$ to
$j_l$ are denoted by $\tilde t_l$ and from $j_r$ to $\tilde j_r$ by
$ \tilde t_r$.\cite{footnotet_lr}  
Proceeding similarly as in Eq.\ (\ref{rechts}) one can express 
$ G_{N,1}= \langle N|G|1\rangle $ as
\begin{equation}
 \label{dot}
G_{N,1}=-G_{N,j_l} \tilde t_l G_{\tilde j_l,1}^{0,l}=
 G_{N,\tilde j_r }^{0,r} \tilde t_r G_{j_r,j_l} 
\tilde t_l G_{\tilde j_l,1}^{0,l},
\end{equation}
where 
\begin{eqnarray*}
G^{0,a}(z)= [z-H_{1,p}(\tilde t_a=0)]^{-1}
\end{eqnarray*}
is the resolvent for the system cut into two pieces by putting $\tilde t_a=0$.
Since $G^{0,a}$ enters the expressions for the
transmission probability derived below, we emphasize that $G^{0,a}$ is 
not the Green function for the Hamiltonian (\ref{Hamiltonian}) but an 
auxiliary Green function.
Using the projection
technique again the problem
of calculating  $G_{j_r,j_l} $ can be reduced to the inversion of an
$N_D\times N_D$ matrix. With $P_{\rm dot}= \sum_{j=j_l}^{j_r}
|j\rangle \langle j| $ one obtains
\begin{equation}
P_{\rm dot} G(z)P_{\rm dot} = \left [ zP_{\rm dot}-
P_{\rm dot} H_sP_{\rm dot}-H_b(z)\right ]^{-1},
\end{equation}
where the boundary term is given by
\begin{equation}
\label{Hb}
H_b(z)=\Gamma_l(z)| j_l\rangle \langle j_l| +
\Gamma_r(z)|j_r\rangle \langle j_r|, 
\end{equation}
with 
\begin{eqnarray*}
\Gamma_a(z)= {\tilde t_a}^2 G_{\tilde j_a,\tilde j_a }^{0,a}(z) \; .
\end{eqnarray*}
As the information from outside the dot is contained in $\Gamma_a$
 it is desirable    
to express the conductance also in terms of these functions.
If one defines the projector 
$
P_l= \sum_{j=1}^{\tilde j_l}
|j\rangle \langle j|  $
and using Eq.~(\ref{projection}) one can derive 
the following identity for $ G_{ P_l P_l}^{0,l} =
 P_l  G^{0,l} P_l $,
\begin{eqnarray}
\label{oth}
&& G_{P_lP_l}^{0,l}(\varepsilon+i0)
- G_{P_lP_l}^{0,l}(\varepsilon-i0) = \nonumber \\
&& -it_L^2\sqrt{4-\varepsilon^2} G_{P_lP_l}^{0,l}(\varepsilon+i0) 
|1\rangle \langle 1| G_{P_lP_l}^{0,l}(\varepsilon-i0).
\end{eqnarray}
 The $\tilde j_l,\tilde j_l$ matrix element of this
 equation relates
$| G_{\tilde j_l,1}^{0,l}|^2$ to the imaginary part of
 $ G_{\tilde j_l,\tilde j_l}^{0,l}$.
With the analogous relation for the right neighbor of the dot one finally
obtains using Eqs.~(\ref{transmission}) and (\ref{dot})
\begin{equation}
|t(\varepsilon)|^2=4 \Delta_l(\varepsilon)\Delta_r(\varepsilon)
 |G_{j_r,j_l}(\varepsilon+i0)|^2.
\end{equation}
Here we have decomposed 
\begin{eqnarray*}
\Gamma_a(\varepsilon+i0)=
\Omega_a(\varepsilon)-i\Delta_a(\varepsilon) \; ,
\end{eqnarray*}
with $\Omega_a $ and $\Delta_a $  real functions. This equation 
is a useful generalization of Eq.\ (\ref{transmission}).

The explicit result for the transmission probability is simplest
for the case $N_D=1$. Then $H_{\rm dot}= P_{\rm dot} H_s  P_{\rm dot}$
consists of a single term\cite{footnoteV_g} 
$\tilde V_g|j_l\rangle \langle j_l|$ 
and one obtains the generalized
Breit-Wigner form\cite{Taylor} 
\begin{equation} 
\label{ND1}
  |t(\varepsilon)|^2 = \frac{4 \Delta_l(\varepsilon)
  \Delta_r(\varepsilon)}{\left[ \varepsilon - \tilde V_g - 
  \Omega_l(\varepsilon) - \Omega_r(\varepsilon)\right]^2 + \left[ 
  \Delta_l(\varepsilon) + \Delta_r(\varepsilon)
\right]^2}~.
\end{equation}
In all cases discussed in this paper a resonance occurs at the 
energy $\varepsilon_R$, where $\varepsilon_R - \tilde V_g - 
  \Omega_l(\varepsilon_R) - \Omega_r(\varepsilon_R)  $
vanishes. Only for a symmetric dot with
$\Delta_l=\Delta_r= \Delta$ the peak value is given by one, i.e.\ perfect
transmission. Later Eq.~(\ref{ND1}) will also be used for the
understanding of the conductance in the presence of a single site
impurity.  

For general dot
size, $G_{j_r,j_l}$ can be expressed in terms
of the $\Gamma_a$ and the resolvent of the isolated dot, 
\begin{eqnarray*}
\bar G^0(z)=
(zP_{\rm dot}-H_{\rm dot})^{-1} \; , 
\end{eqnarray*}
using the special 
form of $H_b$ in Eq. (\ref{Hb})  
\begin{equation}
\label{NDg}
G_{j_r,j_l}=\frac{\bar G_{j_r,j_l}^0}
           {(1-\Gamma_l\bar G_{j_l,j_l}^0)
           (1-\Gamma_r\bar G_{j_r,j_r}^0)
             -\Gamma_l\Gamma_r \bar G_{j_l,j_r}^0\bar G_{j_r,j_l}^0 } .
\end{equation}
For $|\Gamma_l|,|\Gamma_r|\ll 1$
one can have $N_D$ narrow resonances which have the generalized
Breit-Wigner form as in Eq.~(\ref{ND1}). This follows from
Eq.~(\ref{NDg}) by using the spectral representation 
\begin{equation}
\label{Gdot}
\bar G_{j,j'}^0(z)=\sum_{\alpha=1}^{N_D}\frac{\langle
  j|\varepsilon_\alpha\rangle  \langle \varepsilon_\alpha|j'\rangle}
 {z-\varepsilon_\alpha},   
\end{equation} 
where the $|\varepsilon_\alpha\rangle  $ are the eigenstates of $H_{\rm dot}$.
The corresponding eigenvalues $ \varepsilon_\alpha $ depend   linearly
on $\tilde V_g$.
For dots weakly coupled to the rest of the system
the poles in Eq.~(\ref{Gdot}) lead to narrow
resonances in $G_{j_r,j_l} $. For energies close
 to $\varepsilon_\alpha$ the resolvent
matrix elements  $\bar G_{j,j'}^0$ can be replaced by the
corresponding single-pole
 term. Then the terms quadratic in $\bar G^0 $ in the denominator of
Eq.~(\ref{NDg}) cancel and for $\varepsilon\approx
\varepsilon_ \alpha$   one obtains
\begin{eqnarray} 
\label{NDres}
&& \hspace{-.4cm}|t(\varepsilon)|^2 \approx \\
&&\hspace{-.4cm} \frac{4 \Delta_l^{(\alpha)}(\varepsilon)
  \Delta_r^{(\alpha)}(\varepsilon)}{\left[ \varepsilon - \varepsilon_\alpha - 
  \Omega_l^{(\alpha)}(\varepsilon) - \Omega_r^{(\alpha)}
(\varepsilon)\right]^2 + \left[ 
  \Delta_l^{(\alpha)}(\varepsilon) + \Delta_r^{(\alpha)}(\varepsilon)
\right]^2}~, \nonumber
\end{eqnarray}
where $ \Delta_{a}^{(\alpha)}(\varepsilon)=
|\langle j_{a}|\varepsilon_\alpha \rangle |^2
  \Delta_{a}(\varepsilon) $ and correspondingly for
  $\Omega_{a}^{(\alpha)} $. 
The factor
  $|\langle j_{a}|\varepsilon_\alpha \rangle |^2 $ is typically of order
  $1/N_D$ and the widths of the resonances  are reduced by this factor
  compared to the $N_D=1$ dot. As the separation of the resonances is of
order $B/N_D$  the generalized
Breit-Wigner form as in Eq.~(\ref{ND1}) holds for  $|\Gamma_l|,|\Gamma_r|\ll
1$. This can be fulfilled for $\tilde t_l^2, \tilde t_r^2 \ll 1$ 
or for large site impurities 
at sites $\tilde j_l$ and $\tilde j_r$ (see below).
If $\Gamma_a(\varepsilon)$ is assumed to be slowly varying over energies
of the resonance  
 width the integrated weight $w_\alpha$ of the Lorentzian resonance
near $\varepsilon_\alpha$ is given by 
\begin{eqnarray*}
w_\alpha \approx
4\pi [1/\Delta_l^{(\alpha)}(\varepsilon_\alpha)
+1/\Delta_r^{(\alpha)}(\varepsilon_\alpha)]^{-1}
\; . 
\end{eqnarray*}
Here we have assumed that $1-(\Omega_a^{(\alpha)})'  \approx 1$, which
holds for large dots with separated resonances.

Using the general projection formula (\ref{Loewdin}) a relation 
between the diagonal matrix elements of the auxiliary Green function 
$G_{\tilde j_a,\tilde j_a }^{0,a}$ to the left and right of the dot 
[entering Eqs.\ (\ref{ND1}) and (\ref{NDres})] and the corresponding 
matrix elements of the exact Green function  
$G_{\tilde j_a,\tilde j_a }$ can be derived. 
For general dot size and with $\bar a \neq a$ one obtains
\begin{eqnarray}
&& \hspace{-.45cm} G_{\tilde j_a,\tilde j_a }  =  
\frac{\left( 1- \Gamma_{\bar a} \bar G_{j_{\bar a},j_{\bar a}}^0 \right)
G_{\tilde j_a,\tilde j_a }^{0,a}}{\left( 1 -\Gamma_{l} \bar
  G_{j_{l},j_{l}}^0 \right) 
\left( 1 -\Gamma_{r} \bar G_{j_{r},j_{r}}^0 \right) - \Gamma_l
\Gamma_r \bar G_{j_l,j_r}^0\bar G_{j_r,j_l}^0} \nonumber \\*
&& \hspace{-.1cm} 
\approx \frac{\left[ \varepsilon - \varepsilon_{\alpha} - \Omega_{\bar
    a}^{(\alpha)}(\varepsilon) + i \Delta_{\bar
    a}^{(\alpha)}(\varepsilon) \right] G_{\tilde
  j_a,\tilde j_a }^{0,a}(\varepsilon+i0) }{ 
\varepsilon - \varepsilon_{\alpha} -
\Omega_{l}^{(\alpha) }(\varepsilon)- \Omega_{r}^{(\alpha)}(\varepsilon) + i
\left[\Delta_{l}^{(\alpha)}(\varepsilon) +  
\Delta_{r}^{(\alpha)}(\varepsilon)   \right]} \; ,\nonumber 
\end{eqnarray}
 \vspace{-.6cm}
\begin{equation}
\label{Greenrel}
\vspace{-.3cm}
\end{equation}
where the second line holds for $z=\varepsilon+i0 \approx
\varepsilon_{\alpha}$ in the case of dots which are only weakly
coupled to the rest of the system. 

For the interpretation of the temperature dependence of the
conductance of a dot with narrow resonances 
in the regime $  B/N_D \ll T \ll B$ also
 the first equality in Eq. (\ref{dot}) is useful.
 Applying the general
projection formula~(\ref{Loewdin}) yields 
\begin{eqnarray*}
G_{N,j_l}=\frac{ G_{N,j_l}^{0,l}}{1-\tilde t_l^2 G_{\tilde j_l,\tilde
  j_l}^{0,l}G_{j_l,j_l}^{0,l}} \; . 
\end{eqnarray*}
With Eq. (\ref{oth})
and the corresponding relation for $G_{N,j_l}^{0,l}$ one obtains
another exact relation for the transmission probability 
\begin{equation}
\label{single}
|t(\varepsilon)|^2=\frac{4\Delta_l(\varepsilon)\mbox{Im} \,
  G_{j_l,j_l}^{0,l}(\varepsilon+i0)}{\left | 1-\Gamma_l(\varepsilon+i0)
  G_{j_l,j_l}^{0,l}(\varepsilon+i0) \right |^2}.
\end{equation}
If $|\Gamma_l| \ll 1 $ but $|\Gamma_r| \approx 1$ no resonances occur and
perturbation theory in $\Gamma_l$ is possible.
 Putting the denominator in
Eq. (\ref{single}) equal to $1$ leads to the ``golden-rule'' approximation.
 The conductance
for this ``single-impurity case'' is then to leading order in 
$\tilde t_l^2$ given by
\begin{eqnarray}
\label{gTl}
g_l(T)= G_l(T)\frac{h}{e^2} & \approx & 4 \tilde t_l^2 \int_{-B/2}^{B/2}
\left( -\frac{df}{d \varepsilon} \right)  \mbox{Im} \,
  G_{\tilde j_l,\tilde j_l}^{0,l}(\varepsilon+i0) \nonumber \\[.2cm]
&& \times  \mbox{Im} \, G_{j_l,j_l}^{0,l} (\varepsilon+i0) d\varepsilon.
\end{eqnarray}

We now return to the dot situation
with narrow resonances. Its simplest realization is given by 
two hopping impurities  with $\tilde t_l^2, \tilde t_r^2 \ll 1$ 
in an otherwise perfect lattice. Then 
the $G_{\tilde j_a,\tilde j_a}^{0,a}$ and 
$|\langle j_a|\varepsilon_\alpha \rangle |^2$ for $a=l,r$ 
are identical  and the expression for the weights of the 
resonances simplifies to 
\begin{eqnarray*}
w_\alpha \approx 4\pi \, \frac{|\langle j_l|\varepsilon_\alpha \rangle
  |^2 \, 
|\mbox{Im} \, G_{\tilde j_l,\tilde j_l}^{0,l}(\varepsilon_\alpha)|}
{\tilde t_l^{-2}+ \tilde t_r^{-2}} \; .
\end{eqnarray*}
 For    $B/N_D \ll T \ll B$ 
 the transmission probability
$|t(\varepsilon)|^2$
in the Landauer-B\"uttiker
formula (\ref{LB}) can approximately be replaced by $\sum_\alpha w_\alpha
\delta(\varepsilon-\varepsilon_\alpha)$. This yields
\begin{eqnarray}
\label{gTdot}
g_{\rm dot}(T) & \approx & 4 \left[ \frac{1}{\tilde t_l^2}+
  \frac{1}{\tilde t_r^2}\right ]^{-1}
 \int_{-B/2}^{B/2} \left(-\frac{df}{d \varepsilon}\right)
 \mbox{Im} \, G_{\tilde j_l,\tilde j_l }^{0,l}(\varepsilon+i0) \nonumber
 \\[.2cm]  
&& \times  \mbox{Im} \, \bar G_{j_l,j_l}^0(\varepsilon+i0)  d\varepsilon.
\end{eqnarray}
The integrands in  Eqs. (\ref{gTl}) and  (\ref{gTdot}) differ in the
last factor.
For the simple example with $\tilde t_l^2 \ll 1$ in an otherwise 
perfect lattice $ G_{j_l,j_l}^{0,l}$ is just the boundary Green
function $G_b^0$ of a semi-infinite lattice.
For $N_D \gg 1$  and $\tilde t_r^2
 \ll 1$ the imaginary part 
of the boundary Green
 function $ \mbox{Im} \, \bar G_{j_l,j_l}^0 $ of 
 the dot can for the purpose of integration in Eq. (\ref{gTdot}) be
approximately replaced by $ \mbox{Im} \, G_b^0 $.
In Sect.\ \ref{Results} we  encounter more interesting realizations 
for which the
simplified expression for $w_\alpha$ as well as this replacement
can be used.
 Comparison 
of Eqs. (\ref{gTl}) and  (\ref{gTdot}) then shows that
in the ``high-temperature regime''  $B/N_D \ll T \ll B$ 
the conductance of the dot
can be obtained by adding resistances for the single-impurity
problems (Kirchhoff's law) described by Eq. (\ref{gTl}),
\begin{equation}
\label{Kirchhoff}
\frac{1}{g_{\rm dot}(T)}\approx \frac{1}{g_l(T)}+\frac{1}{g_r(T)}.
\end{equation}
The quantum mechanical interference effects in the dot are  washed out
for large enough temperatures.
For lower temperatures the approximation Eq. (\ref{Kirchhoff}) breaks down.
 Another temperature
window where the result for the conductance takes a simple form
is given by $|\Delta_a^{(\alpha)}(\varepsilon_\alpha)|\ll T \ll B/N_D$.
If the resonance is at the chemical potential or much closer to  it
than $T$ the
conductance decays with temperature $\propto w_\alpha/T$. For the
interacting case discussed in 
Sect.\ \ref{Results} the weight of the resonance $ w_\alpha $
itself is temperature dependent which results in $G(T)\propto
T^{\alpha_B-1}$.

The above arguments for weak hopping connections of the dot
with the rest of the system can easily be generalized to the
case of $\tilde t_a \approx 1 $ but a large site impurity
  $V_{a} $  at the site
$\tilde  j_a$. For $|V_l|\gg 1 $ the site $\tilde  j_l $ can
be ``integrated out'' using Eq.\ (\ref{Loewdin}). This leads to a weak hopping 
$\tilde t_{j_{l-2},j_l}=\tilde t_{j_{l-2},j_{l-1}} \, 
\tilde t_l /  V_l $ between the sites $j_{l-2}$ and $j_l$ .

The relations presented in this section play an
important role for the
interpretation of our fRG results for the transport through interacting
quantum wires.

\section{The functional renormalization group at $\bm{T>0}$}
\label{fRGintransport}

We now return to our Hamiltonian Eq.\ (\ref{fullH}). 
Due to the presence of leads the direct 
calculation  of 
the non-interacting  
propagator related to $H_{\rm kin} + H_{\rm imp}$ [see Eqs.\ (\ref{kindef})
and (\ref{impdef})] 
requires the inversion of an infinite matrix. Using 
Eq.\ (\ref{Loewdin}) it can be reduced to the inversion 
of an $N \times N$ matrix. 
The leads then provide an additional diagonal one-particle potential
on sites $1$ and $N$, depending on the Matsubara frequency $\omega_n$,
\begin{eqnarray}
\label{leadpotdef}
 V_{j,j'}^{\rm lead}(i\omega_n) & = &
 \frac{i\omega_n+\mu}{2} \left( 1 - 
 \sqrt{1 - \frac{4}{(i\omega_n+\mu)^2}} \, \right) \nonumber \\
&& \times  \delta_{j,j'} \left( \delta_{1,j} + \delta_{N,j} \right) \; .
\end{eqnarray}
Since the interaction is only non-vanishing
on the bonds between the sites $1$ to $N$ the problem including the
semi-infinite leads is this way reduced to the problem of an 
$N$-site chain. 
The chemical potential of the entire system in thermal equilibrium 
is denoted by $\mu(n,U,T)$. 

To set up the fRG  the projected non-interacting propagator 
$G_0$ is replaced by 
\begin{equation}
\label{cutoffproc}
G_0^{\Lambda}(i \omega_n) = \chi^{\Lambda}(\omega_n)
G_0(i \omega_n)
\end{equation}
with a function $\chi^{\Lambda}$ which is 
unity for $|\omega_n| \gg \Lambda$ and vanishes for 
$|\omega_n| \ll \Lambda$. At $T=0$ we earlier considered 
$\chi^{\Lambda}(\omega) = \Theta(|\omega|-\Lambda)$.\cite{VM1,VM2,Sabine} 
For technical reasons at $T>0$ a smoother cutoff function, 
specified below, is more appropriate. We here choose to include the impurity 
part of the Hamiltonian $H_{\rm imp}$ in $G^0$. Earlier it was taken 
as a part of the self-energy.\cite{Sabine} 
At $T=0$ both choices lead to the same results. 
By means of $G_0^{\Lambda}$  the generating functional 
of the one-particle irreducible vertex functions 
for the Hamiltonian $H$ 
becomes $\Lambda$ dependent. 
Differentiating the functional with respect to $\Lambda$
and expanding it 
in the external sources then leads to an exact infinite hierarchy
of coupled flow equations for the vertex 
functions.\cite{Wetterich,Morris,SalmhoferHonerkamp}  
In practical applications this set of equations has to be truncated.
As a first step we neglect the three-particle vertex
which is small as long as the two-particle vertex $\Gamma^{\Lambda}$ 
stays small which is the case for not-too-large $U$.\cite{Sabine} 
This approximation leads to a closed set of equations for 
$\Gamma^{\Lambda}$  and the self-energy $\Sigma^{\Lambda}$ 
given by
\begin{equation}
\label{flowS}
 \frac{\partial}{\partial\Lambda} \Sigma^{\Lambda}(1',1) =
 \, - \, \frac{1}{\beta} \, \sum_{2,2'} \, e^{i \omega_{n_2} 0^+} \,
 S^{\Lambda}(2,2') \; \Gamma^{\Lambda}(1',2';1,2) 
\end{equation}
and
\begin{eqnarray}
 && \frac{\partial}{\partial\Lambda} \Gamma^{\Lambda}(1',2';1,2)
 = \; \frac{1}{\beta} \,
 \sum_{3,3'} \sum_{4,4'} \, G^{\Lambda}(3,3') \, S^{\Lambda}(4,4')
 \nonumber \\
 && \times \Big[
 \Gamma^{\Lambda}(1',2';3,4) \, \Gamma^{\Lambda}(3',4';1,2) 
 \nonumber \\
&&-  \Gamma^{\Lambda}(1',4';1,3) \, \Gamma^{\Lambda}(3',2';4,2) 
 - (3 \leftrightarrow 4, 3' \leftrightarrow 4') \nonumber \\
 &&+  \Gamma^{\Lambda}(2',4';1,3) \, \Gamma^{\Lambda}(3',1';4,2)
 + (3  \leftrightarrow 4, 3'  \leftrightarrow 4') 
 \, \Big] \nonumber \; .
\end{eqnarray}
\vspace{-.7cm}
\begin{equation}
\label{flowGamma}
\vspace{-.2cm}
\end{equation}
The labels $1$, $2$, etc., stand for the quantum numbers of the
non-interacting one-particle basis and the Matsubara
frequencies. 
The full propagator $G^{\Lambda}$ 
is given by
\begin{equation}
\label{Gdef}
 G^{\Lambda} = \left[ (G_0^{\Lambda})^{-1} - \Sigma^{\Lambda} \right]^{-1}
\end{equation}
and the so-called single-scale propagator $S^{\Lambda}$ by 
\begin{eqnarray}
\label{singlescale}
 S^{\Lambda}  =
 G^{\Lambda} \left[ \frac{\partial}{\partial \Lambda} 
\left(G_0^{\Lambda}\right)^{-1} \right] 
 G^{\Lambda}  \; . \\ \nonumber
\end{eqnarray}

\begin{figure}[htb]
\begin{center}
\includegraphics[width=0.49\textwidth,clip]{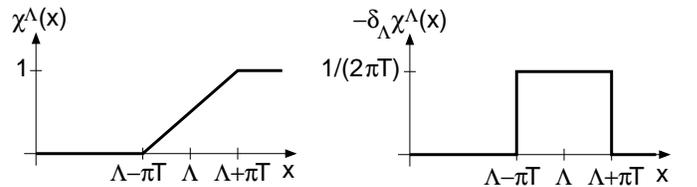}
\end{center}
\vspace{-0.3cm}
\caption[]{The cutoff function $ \chi^\Lambda(x)$ for fixed $\Lambda$
and the negative of its derivative (with respect to 
$\Lambda$) . \label{tcutoff}}
\end{figure}

We choose the cutoff function 
\begin{eqnarray*}
  \chi^\Lambda(\omega_n) & :=
 \left\{  \begin{array}{ll}  
    0  \, ,& |\omega_n| \leq \Lambda-  \pi T \\
    \frac{1}{2} + \frac{|\omega_n|-\Lambda}{2\pi T}
    \, , & \Lambda - \pi T \leq |\omega_n | \leq  \Lambda +\pi T \\
    1  \, ,& |\omega_n| \geq  \Lambda  + \pi T
  \end{array} \right.
\end{eqnarray*}
with the accompanying derivative
\begin{eqnarray*}
  -  \frac{\partial}{\partial\Lambda} \chi^\Lambda(\omega_n) & =
 \left\{  \begin{array}{ll}
    \frac{1}{2\pi T} \, ,
    &\Lambda - \pi T \leq |\omega_n | \leq  \Lambda +\pi T \\
    0  \, ,& \text{otherwise} \, ,
  \end{array} \right.
\end{eqnarray*}
where $\Lambda$ starts at $\infty$ and goes down to $0$. 
Both these functions are shown in Fig.~\ref{tcutoff}.
With this choice and for a fixed $\Lambda$ 
a Matsubara sum involving the single-scale propagator
Eq.\ (\ref{singlescale}) contains only the two terms with  
frequencies $\Lambda-\pi T < |\omega_n| < \Lambda+\pi T$.
This reduces the numerical effort in solving the coupled 
differential equations for the self-energy and the two-particle 
vertex.
With this cutoff function the right-hand side of the flow equation has
a jump at every Matsubara frequency and is smooth in between, such
that the flow can be integrated by standard routines.  

To obtain a manageable set of equations even for large $N$ 
we next introduce further approximations in the flow equation of
the two-particle vertex.\cite{Sabine} 
We are dealing with at most two local impurities. For large $N$ such a
small number of impurities does not significantly alter the flow of 
the effective interaction and we thus neglect the feedback of the 
bare impurity potential 
entering $G_0^{\Lambda}$ on the flow of $\Gamma^{\Lambda}$.
Self-energy effects on the flow of 
$\Gamma^{\Lambda}$ are of order $U^3$ and are ignored.  
For the integration of Eq.\ (\ref{flowGamma}) we furthermore consider a 
system with constant $U$ on all bonds. The two-particle vertex is projected
onto the Fermi points and parameterized
by a nearest-neighbor interaction $U^{\Lambda}$. 
After taking the limit $N \to \infty$ 
the flow equation for $U^{\Lambda}$ reads\cite{Sabine}
\begin{eqnarray} 
\label{flowU}
\frac{\partial}{\partial \Lambda} 
U^\Lambda = h(\tilde\omega_n) \, (U^\Lambda)^2
\end{eqnarray}
with 
\begin{widetext}
\begin{eqnarray}
 h(x) = - \frac{1}{2\pi} -
 \mbox{Re} \, \bigg[ \frac{i}{2} \, (\mu_0 + i x) 
 \, \sqrt{1 - \frac{4}{(\mu_0 + ix)^2}} \; \frac
 {3i\mu_0^4 - 10\mu_0^3 x - 12i\mu_0^2(x^2 + 1) + 6 x^3\mu_0
  + 18 x \mu_0 + 6i x^2 + i x^4}
 {\pi (2\mu_0 + ix)(4 - \mu_0^2 + x^2 - 2i x \mu_0)^2} \,
  \bigg] , 
\end{eqnarray}
\end{widetext}
where $\tilde\omega_n$ stands for the fermionic Matsubara frequency
closest to $\Lambda$, and $\mu_0 = -2\cos k_F$. 
The above simplified flow equation yields not only the correct 
low energy asymptotics to second order in the renormalized vertex,
but contains also all non-universal second order corrections 
to the vertex at $\pm k_F$ at higher energy scales.\cite{Sabine}
In the case where the 
interaction depends on position,
as an additional approximation we apply Eq.\ (\ref{flowU}) locally 
for each bond. As long as the bulk part of the interacting wire is
much larger than the contact regions this has a small effect.  

Because of the above approximations the self-energy is a
frequency-independent tridiagonal matrix in real space: only
$\Sigma_{j,j}$ and $ \Sigma_{j,j\pm 1}$ are non-zero.  
In the exact solution a frequency dependence of $\Sigma$ 
is generated to order $U^2$ (bulk TLL behavior), which
shows that in the approximation for the self-energy terms of 
order $U^2$ are only partly included. For the flow of
the $T$- and $N$-dependent self-energy we obtain
\begin{eqnarray}
 \frac{\partial}{\partial \Lambda} \Sigma_{j,j}^\Lambda  = 
    -\frac{1}{2\pi}
    \sum_{|\omega_n| \approx \Lambda} \sum_{r=\pm 1}
    U_{j,j+r}^\Lambda \left[ \frac{1}{Q(i\omega_n)-\chi^\Lambda(\omega_n)
       \Sigma^\Lambda} \right. \nonumber \\*  \left. \times    
      Q(i\omega_n) \, \frac{1}{Q(i\omega_n)-\chi^\Lambda(\omega_n)
      \Sigma^\Lambda } \right]_{j+r,j+r} \nonumber\\*
 \frac{\partial}{\partial \Lambda}     \Sigma_{j,j\pm 1}^\Lambda  = 
    \frac{1}{2\pi}
    \sum_{|\omega_n| \approx \Lambda} U_{j,j\pm 1}^\Lambda 
\left[ \frac{1}{Q(i\omega_n)-\chi^\Lambda(\omega_n) \Sigma^\Lambda }
  \right.  \nonumber \\* 
\left. \times Q(i\omega_n) \, \frac{1}{Q(i\omega_n)-\chi^\Lambda(\omega_n)
\Sigma^\Lambda} \right]_{j,j\pm 1} \; , \nonumber
\end{eqnarray}
\vspace{-.5cm}
\begin{equation}
\vspace{-.2cm}
\label{eq:sigmaflowtg0}
\end{equation} 
where $|\omega_n| \approx \Lambda$ stands for taking the positive 
as well as negative frequency with absolute value closest to $\Lambda$,
and $j,j\pm 1 \in [1,N]$.  
The matrix $Q$ is the inverse of the cutoff-independent, projected, 
and non-interacting propagator $Q=(G_0)^{-1}$. 

The initial condition of the self-energy at $\Lambda=\infty$ is given by 
\begin{eqnarray*}
\Sigma^{\infty}_{j,j} & = & - \nu(n,U) \left( U_{j-1,j} +
  U_{j,j+1} \right) \\  
\Sigma^{\infty}_{j,j\pm 1} & = & 0 \; ,
\end{eqnarray*}
and the one of the flowing nearest-neighbor interaction by $U^{\infty}
=U$. In a numerical solution the flow starts at a large finite initial
cutoff $\Lambda_0$. One has to take into account that, due to 
the slow decay of the right-hand side of the flow equation 
for $\Sigma^{\Lambda}$, the integration from $\Lambda=\infty$ to
$\Lambda=\Lambda_0$ yields a contribution which does not vanish for
$\Lambda_0 \to \infty$, but rather tends to a finite
constant.\cite{Sabine} The resulting initial condition at 
$\Lambda = \Lambda_0 \to \infty$ reads
\begin{eqnarray*}
\Sigma^{\Lambda_0}_{j,j} & = & 
\left[1/2- \nu(n,U)\right] \left( U_{j-1,j} +
  U_{j,j+1} \right)  \\ \Sigma^{\Lambda_0}_{j,j\pm 1}  & = & 0 \; .
\end{eqnarray*}
The right-hand side of Eq.~(\ref{flowU}) tends to zero fast enough so
that no such additional contribution is built up for 
the flowing interaction and $U^{\Lambda_0} = U$.

Neglecting the leads, at $T=0$, and in the limit of a single weak 
or strong impurity the coupled differential equations (\ref{flowU}) 
and (\ref{eq:sigmaflowtg0}) can be solved analytically.\cite{VM2} 
The results are consistent with what is expected from the LSGM.
For general parameters we  
solve the flow equations numerically. We have developed an
algorithm\cite{footnotealgo} which for $T>0$ approximately scales with 
$N/T$. For finite $T$ systems of $10^4$ lattice sites are 
considered, roughly corresponding to the length of quantum wires 
accessible to transport experiments.  
Since the interacting wire consists of $N$ lattice sites the energy 
scale 
\begin{eqnarray}
\label{levelspacing}
\delta_N = \frac{\pi v_F}{N} 
\end{eqnarray}
forms a lower bound for any temperature scaling of the conductance 
with an interaction dependent exponent
(see Sect.\ \ref{Results}) and $G(T)$ quickly saturates 
for $T \lessapprox \delta_N$.
Therefore only temperatures of the order of the band width down 
to $T=10^{-4}$ are relevant and have been studied. The Fermi velocity
is given by $v_F=2 \sin k_F$, with the Fermi momentum $k_F=n \pi$.  
We note that the saturation of
$\Sigma^{\Lambda}$ for $\Lambda \lessapprox T$ or $\Lambda \lessapprox 
\delta_N$
sets in ``automatically'', in contrast
to more intuitive renormalization group schemes in which the flow 
of the considered quantities is stopped ``by hand'' by replacing 
$\Lambda \to T$ or $\Lambda \to \delta_N$, 
respectively.\cite{KaneFisher,Furusaki0,MatveevGlazman,Nazarov,Polyakov}  

At the end of the flow ($\Lambda=0$),
$\Sigma^{\Lambda=0}$ presents an approximation for 
the exact self-energy and will be denoted by $\Sigma$ in what follows.
The fRG is set up in the grand canonical ensemble with fixed chemical
potential $\mu$. As we want to compare our results with those
obtained for the LSGM considering the canonical ensemble with fixed
density $n$ we first determine the parameter $\nu(n,U)$ such that 
at $T=0$ the density on the sites $1$ to $N$ acquires 
the desired value $n$. In a second step the global chemical potential 
$\mu(n,U,T)$ is determined such that the density of the entire 
system (interacting wire and leads) remains $n$ for all $T$. Due to 
particle-hole symmetry in the half-filled case we have $\nu(1/2,U)=1/2$ 
and $\mu(1/2,U,T)=0$. 

\begin{figure}[htb]
\begin{center}
\includegraphics[width=0.49\textwidth,clip]{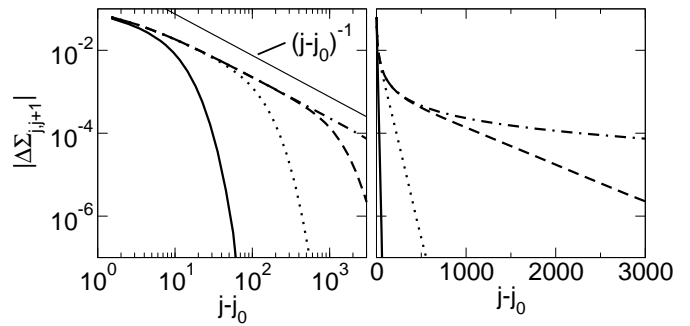}
\end{center}
\vspace{-0.3cm}
\caption[]{ Decay of the oscillatory part of the off-diagonal matrix
  element of the self-energy away from a single hopping impurity at
  bond $j_0,j_0+1$. Results for $t'=0.1$, $j_0=5000$, $N=10^4$, $U=1$,
  $n=1/2$ and different temperatures $T=10^{-1}$ (solid line),
  $T=10^{-2}$ (dotted line), $T=10^{-3}$ (dashed line), and
  $T=10^{-4}$ (dashed-dotted line) are presented.  The left panel
  shows the data on a log-log scale, the right panel on a linear-log scale.
  For comparison the left panel contains a power-law $(j-j_0)^{-1}$
  (thin solid line). \label{self}}
\end{figure}

Typical results of the $N$-dependent $\Sigma_{j,j}$ and 
$ \Sigma_{j,j+ 1}$ at $T=0$ for a single impurity located at site 
$j_0$ and $U>0$ were presented in Figs.\ 5 and 6 of 
Ref.~\onlinecite{Sabine}. 
Both matrix elements show long-ranged spatial oscillations induced 
by the impurity with an amplitude which asymptotically decays as 
$1/|j-j_0|$.\cite{Sabine}
For finite $T$ this power-law decay of the self-energy induced by a 
single impurity is cut off at a scale $\propto 1/T$ beyond which 
the oscillatory part of 
$\Sigma_{j,j}$ and $ \Sigma_{j,j+ 1}$ decays exponentially. This is 
exemplified in Fig.~\ref{self} in which 
$\left| \Delta\Sigma_{j,j+ 1} \right| = 
\left| \Sigma_{j,j+ 1} - \bar \Sigma_{o} \right|$ is shown 
for different $T$ and as a function of $j-j_0$. The average value of 
$\Sigma_{j,j+1}$ away from the impurity
site, which is related to the interaction-induced broadening of the band,
is denoted by $ \bar \Sigma_{o}$. In the left panel of
Fig.~\ref{self}, $\left| \Delta\Sigma_{j,j+ 1} \right|$ is shown on a
log-log scale where straight lines are power-laws, while in the right
panel it is presented on a linear-log scale where straight lines
represent an exponential decay.
The self-energy was calculated for 
the case of a single hopping impurity with $t'=0.1$ located
in the middle of the interacting wire with $N=10^4$ sites.  
The other 
parameters are $U=1$ and $n=1/2$. For a hopping impurity and at 
half filling the diagonal part of the self-energy $\Sigma_{j,j} $ 
vanishes. 

From $\Sigma$ we obtain an approximation for
the conductance $G(T)$ through our interacting system. 
We use the fact that it is consistent to neglect vertex corrections 
to the Landauer-B\"uttiker formula when the approximate self-energy is
frequency independent.\cite{Oguri} This is the case in our fRG scheme 
as discussed in the Appendix. 
Therefore using Eq.\ (\ref{LB}) does not 
imply any further approximation. To obtain $G(T)$ we have to calculate 
the effective, $T$- and $\delta_N$-dependent transmission 
$|t(\varepsilon,T,\delta_N)|^2$ (where the dependences on 
$\varepsilon$, $T$, and $\delta_N$ are from now on written
explicitly)
which is expressed in terms of the one-particle Green function  
between sites 1 and $N$ [see Eq.\ (\ref{transmission})] and can
be calculated by inverting the tridiagonal matrix $Q-\Sigma$. 
This way we have reduced the many-body problem to
a single-particle scattering problem off an effective impurity potential 
on the sites 1 to $N$ given by the 
$T$- and $\delta_N$-dependent self-energy. For the detailed understanding 
of our findings for the single- and double-barrier case we use 
the one-particle scattering theory 
results derived in Sect.\ \ref{Scattheory}. 

A comparison of the fRG results for a single impurity 
at $T=0$ to accurate DMRG results and exact scaling properties shows 
that the error due to our various approximations is small for 
interactions such that $1/2 \leq K \leq 1$.\cite{Sabine,VM4} As we 
discuss next, the same holds at finite $T$.

\section{Results}
\label{Results}

\subsection{Single impurity}
\label{Singimp}

In this section we report our results for transport through a single
impurity, and later through a double barrier.  In the single impurity
case we mainly consider site impurities; we have obtained similar 
results for hopping impurities. We first consider impurities placed 
sufficiently away from the contact regions and then briefly discuss
the dependence of the conductance on the impurity position $j_0$. 

\begin{figure}[htb]
\begin{center}
\includegraphics[width=0.4\textwidth,clip]{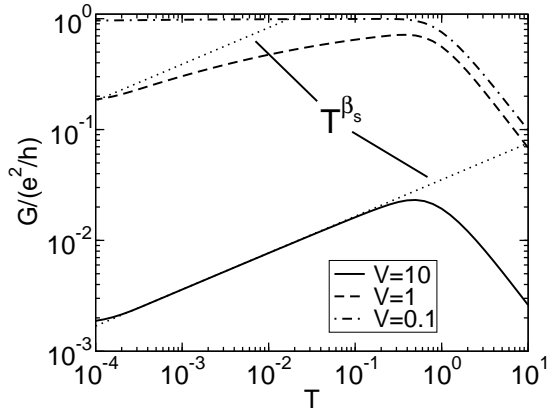}
\end{center}
\vspace{-0.3cm}
\caption[]{Temperature dependence of the conductance for a single
  site impurity with different $V$ as indicated in the legend 
  and $U=0.5$, $n=1/2$, $N=10^4$. The dotted line shows a
  power-law with an exponent $\beta_s$ obtained from fitting the
  $V=10$ data. \label{single1}}
\end{figure}

Fig.~\ref{single1} shows typical fRG results for the $T$ dependence of the 
conductance in the case of a single site impurity of strength $V$. 
The parameters are $U=0.5$, $n=1/2$, and $N=10^4$. 
For $T \geq B$ all curves
show a $1/T$ scaling. For these
temperatures the derivative of the Fermi function in Eq.~(\ref{LB})
only varies very weakly with $\varepsilon$ for the relevant 
energies $-B/2 \leq \varepsilon \leq B/2$ but decreases with increasing 
temperature as $1/T$. At the same time the integral over the band
energies of $|t(\varepsilon,T,\delta_N)|^2$ is only weakly 
$T$-dependent leading to
the observed $1/T$ behavior. For a strong impurity $V=10$ and 
$\delta_N \leq  T \ll B$, $G(T)$ nicely follows a
power-law with an exponent $\beta_s(U,n)$ indicated in
Fig.~\ref{single1} by the dotted line. More specifically we find 
for all $U$ and $n$ that power-law scaling does not set in before 
$T \lessapprox B/40$.
It holds down to 
$T \approx \delta_N/2$ ($\approx 3 \cdot 10^{-4}$ for $N=10^4$) 
beyond which the conductance 
saturates. The exact value of this lower bound depends on the impurity
position $j_0$ (see below).\cite{FuNa} We here consider $j_0 \approx
N/2$. For an intermediate impurity $V=1$ 
the slope of the data (on a log-log scale) tends towards 
$\beta_s$ but is still significantly away from it when saturation 
sets in at $\delta_N/2$.
The slow change of the slope seen in
Fig.~\ref{single1} is a general feature of intermediate $V$. 
This observation is of relevance for the analysis of noisy
experimental data usually restricted to temperatures within one to two
orders of magnitude.\cite{Bockrath,Yao} Under these conditions 
a transient regime might easily incorrectly be identified as the 
asymptotic power-law regime leading to an  
exponent which is too small. For a weak impurity $G(T)$ stays very
close to $e^2/h$  for all $T \ll B$. 

\begin{figure}[htb]
\begin{center}
\includegraphics[width=0.4\textwidth,clip]{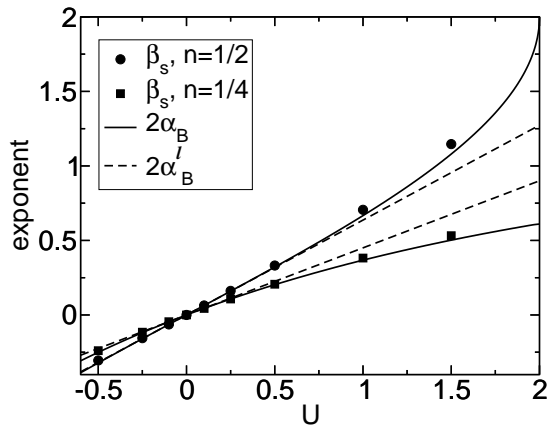}
\end{center}
\vspace{-0.3cm}
\caption[]{Dependence of the strong-impurity exponent $\beta_s$ on 
the interaction $U$. For comparison the LSGM
prediction $2 \alpha_B$ (solid line) and the leading order (in $U$) 
result $2 \alpha_B^l$ (dashed line) are shown.   \label{single2}}
\end{figure}

In Fig.~\ref{single2} the $U$-dependence of the strong-impurity 
exponent $\beta_s$ is shown for $n=1/2$ (circles) and $n=1/4$ (squares). 
We also present results for $U<0$, where $G(T)$ increases 
for decreasing $T$ and $\beta_s<0$. In this case $G(T) \propto
T^{\beta_s}$ does not present the
asymptotic low-energy behavior since for $T \to 0$, $G(T)$ is tending
towards $e^2/h$. The larger $|U|$, the faster $G(T)$
increases. At the same time the temperature regime over which a
power-law exponent can be read off shrinks even if $V$ is
chosen to be very large. 
An exponent can be reliably read off for $U \geq
-0.5$ only. From the studies of the LSGM\cite{KaneFisher,Fendley} 
we expect that  
the strong-impurity exponent is $2\alpha_B=2(1/K-1)$. For $n=1/2$, $K$ is
given in Eq.~(\ref{BetheAnsatz}) while away from half filling it can
be calculated solving the Bethe ansatz integral equations
numerically.\cite{Haldane} For comparison, $2\alpha_B$ (solid line) 
and the leading-$U$ behavior  
\begin{eqnarray}
\label{alpha_B_leading_U}
2 \alpha_B^l = 2 \; \frac{2 U (1-\cos[2 k_F])}{2 \pi v_F} 
\end{eqnarray}
(dashed line) are presented in 
Fig.~\ref{single2}. The exponent $\beta_s$ turns out to be at least 
correct to leading order in $U$. Furthermore, even for intermediate 
$U \lessapprox 1.5$, it is still very close to $2 \alpha_B$ in 
contrast to the leading-order contribution $2 \alpha_B^l$.

\begin{figure}[htb]
\begin{center}
\includegraphics[width=0.4\textwidth,clip]{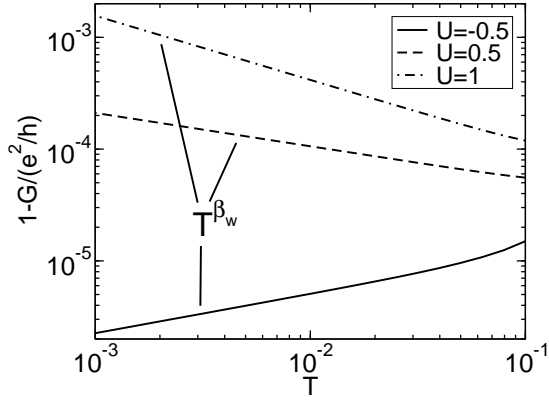}
\end{center}
\vspace{-0.3cm}
\caption[]{Weak-impurity behavior  of $1-G(T)/(e^2/h)$ for 
$V=0.01$, $n=1/2$, $N=10^4$
and different $U$ as indicated in the legend. \label{single3}}
\end{figure}

\begin{figure}[htb]
\begin{center}
\includegraphics[width=0.4\textwidth,clip]{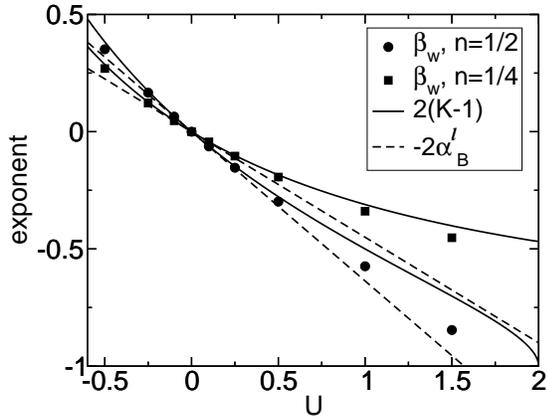}
\end{center}
\vspace{-0.3cm}
\caption[]{Dependence of the weak-impurity exponent $\beta_w$ on 
the interaction $U$. For comparison the LSGM
prediction $2 (K-1)$ (solid line) and the leading order (in $U$) 
result $-2 \alpha_B^l$ (dashed line) are shown.   \label{single4}}
\end{figure}

For weak impurities $G(T)$ is predicted to follow a power-law scaling
$1-G(T)/(e^2/h) \propto T^{2(K-1)}$ which holds as long as the
correction to perfect conductance stays small.\cite{KaneFisher,Fendley} 
For $K-1 > 0$, i.e.\ $U<0$, this scaling presents the asymptotic low-energy 
behavior. In Fig.~\ref{single3},
$1-G(T)/(e^2/h)$ is shown for $V=0.01$, $n=1/2$, $N=10^4$,
and different $U$. We can read off an exponent $\beta_w(U,n)$. In
Fig.~\ref{single4}, $\beta_w$ for $n=1/2$ and $n=1/4$ is compared 
to $2(K-1)$ and the  leading-order term $-2 \alpha_B^l$. Note that to leading 
order $2\alpha_B = 2(1-K)$, which will become important in the 
comparison of the fRG to the ``leading-log'' method. Our weak-impurity
exponent $\beta_w$ agrees to 
first order in $U$ with $2(K-1)$, but is closer to the exact result than 
$-2 \alpha_B^l$. 

\begin{figure}[htb]
\begin{center}
\includegraphics[width=0.4\textwidth,clip]{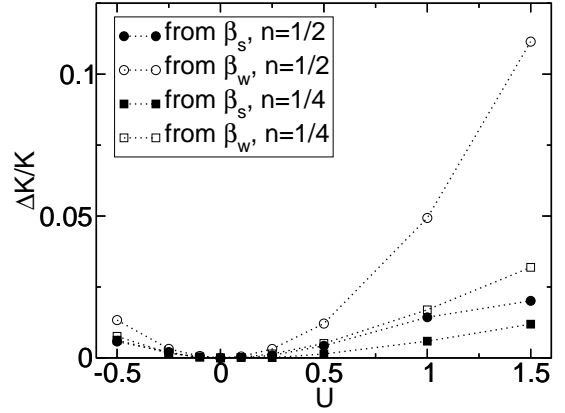}
\end{center}
\vspace{-0.3cm}
\caption[]{Relative error of the fRG approximations for the TLL
  parameter from the strong- $\beta_s$ and weak- $\beta_w$
impurity exponents at $n=1/2$ and $n=1/4$. \label{single5}}
\end{figure}

In the solution of the LSGM the strong- and weak-impurity exponents 
characterizing the scaling of $G$ can both be expressed in terms of
$K$. Identifying $\beta_s=2(1/K_s -1)$ and $\beta_w=2(K_w-1)$ we
obtain two fRG approximations $K_s$ and $K_w$ for $K$.
In Fig.~\ref{single5} the relative error of the fRG approximations
to the exact TLL parameter $\Delta K/K=\left|K-K_{s/w}\right|/K$ is presented
for $n=1/2$ and $n=1/4$. 
From the figure it is apparent that 
$\Delta K \propto U^2$. In our approximation scheme terms of order
$U^2$ are only partially included\cite{Sabine} and we thus cannot expect
agreement to higher order. 
The error $\Delta K/K$ depends on $n$. 
The half-filled case exhibits the largest 
deviation from the exact result.
This can be understood from the flow of $U^{\Lambda}$. For fixed $U>0$
as a function of $n$, $(U^{\Lambda=0} - U)/U$ becomes maximal at $n=1/2$,
where the nearest-neighbor interaction increases by a few 10 percent
(for an example at $T=0$ and $U=1$ see Fig.~4 of Ref.~\onlinecite{Sabine}).
In contrast for smaller fillings, e.g.\ $n=1/4$, $U^{\Lambda}$
even decreases during the flow. This explains why our weak coupling
method works particularly well for small fillings.  
The two approximations $K_s$
and $ K_w$ differ to order $U^2$ which shows that within 
our approach the TLL relation between the strong- and weak-impurity 
exponents is only fulfilled approximately.    

\begin{figure}[htb]
\begin{center}
\includegraphics[width=0.4\textwidth,clip]{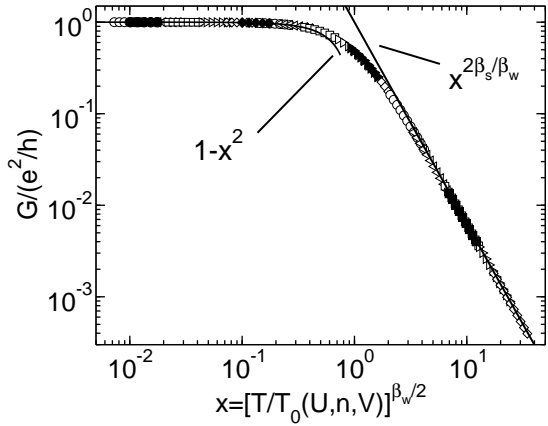}
\end{center}
\vspace{-0.3cm}
\caption[]{One-parameter scaling plot of the conductance. Open
  symbols represent results obtained for $U=0.5$, $n=1/2$, and
  different $T$ and $V$, while filled symbols were calculated for 
  $U=0.851$, $n=1/4$. Both pairs of $U$ and $n$ lead to the same
  $K_w=0.85$. The solid lines indicate the asymptotic behavior for
  small and large $x$.  
 \label{single6}}
\end{figure}

In the LSGM with repulsive interaction even a weak bare impurity 
induces a power-law suppression of $G$ with exponent
$2 \alpha_B$ in the limit where all 
energy scales are sent to zero.
For weak bare impurities and  $1/2 \leq K <1$, 
the crossover scale  
between weak- and strong-impurity behavior goes as $V^{1/(1-K)}$ and 
becomes very small.\cite{KaneFisher,Moon,Fendley}
For the lattice model considered here we have observed 
a similarly small scale.\cite{VM4}              
As a consequence, even for the fairly large system size of 
$N=10^4$ sites and the large range of temperatures we can treat,
it is impossible to directly demonstrate the full 
crossover for a single set of parameters. In the LSGM it can be 
shown indirectly using a one-parameter scaling ansatz 
\begin{eqnarray}
\label{oneparascaling}
G = \frac{e^2}{h} \tilde G_K(x) \; , \;\;\;  
x=\left[ T/T_0(U,n,V)\right]^{K-1} \, 
\end{eqnarray}
with a non-universal scale 
$T_0(U,n,V)$.\cite{KaneFisher,Moon,Fendley} 
For appropriately
chosen $T_0$ the $G(T)$ curves for different $V$ (but fixed 
$K$) can be collapsed onto the $K$-dependent scaling function $\tilde
G_K(x)$. It has the limiting behavior $\tilde G_K(x) \propto 1-x^2$ for $x
\to 0$ and $\tilde G_K(x) \propto x^{-2/K}$ for $x \to \infty$. 
We earlier demonstrated one-parameter scaling for the microscopic lattice 
model considered within our approximation scheme replacing $T$ in 
the above ansatz Eq.\ (\ref{oneparascaling}) by $\delta_N$.\cite{VM4} 
To set up the scaling we have to decide whether we
take $K_s$ or $K_w$ in the definition of the
variable $x$. For a comparison to the
analytically known $K=1/2$ scaling function of the LSGM it is 
advantageous to use $K_s$, since for the corresponding $U$ 
[$U = 2$ for $n=1/2$; see Eq.~(\ref{BetheAnsatz})] 
$K_s$ is much closer to the exact $K$ than $K_w$.
(see Fig.~\ref{single5}). This is what we did in Ref.\
\onlinecite{VM4}. This choice leads to  a small deviation from the $1-x^2$ 
behavior at small $x$. For small $U$ the difference in the 
approximate TLL parameters is small and we here use $K_w$ instead.
In the limit $x \to 0$ it implies an exact $1-x^2$ behavior of the fRG 
approximation for the scaling function. A one-parameter scaling plot
is shown in Fig.~\ref{single6}. The different 
symbols stand for different $V \in [0.01,30]$. For each $V$, $G(T)$
was calculated for several $T\in [10^{-3},10^{-1}]$.  
The open symbols represent results obtained at half filling for $U=0.5$ 
which gives $K_w=0.85$. The same $K_w$ we find for quarter-filling and
$U=0.851$. Data for the latter parameters are shown as filled
symbols. The collapse of the results for $n=1/2$ and $n=1/4$
exemplifies that the scaling function depends on $U$ and $n$ only 
via the TLL parameter $K$ as predicted for the LSGM.   
It is very remarkable that our exponent $2\beta_s/\beta_w$  
of the large $x$-behavior agrees to leading order with the 
corresponding exponent $-2/K$ of the LSGM. This is a significantly stronger
result than the observation that $\beta_s$ and $2 \alpha_B$ as well as 
$\beta_w$ and $2(K-1)$ agree to order $U$.
Considering site (hopping) impurities which are extended over more
than one site (bond) as well as impurities which are given by a
combination of modified on-site energies and hopping matrix elements 
we have verified that $\tilde G_K$ is also independent of 
the details of the local impurity.

A comparison of the results presented here for fixed $\delta_N$ and varying 
$T$ and the results of Ref.\ \onlinecite{VM4} obtained for 
$T=0$ and different $\delta_N$ shows that in the single-impurity case
and for smooth contacts $\delta_N$ and $T$ present equivalent scaling
variables. In alternative RG schemes 
this equivalence is instead assumed. 
Note that the equivalence no longer holds if either non-perfect 
contacts\cite{VM4,footnotenonperfect} or resonant tunneling is 
considered (for the latter, see below).  

In Ref.\ \onlinecite{MatveevGlazman} an alternative fermionic RG method 
was introduced to study an effective low-energy model with a single 
barrier in the limit of weak interaction. In this approach an RG 
equation is set up to resum the ``leading-log'' divergences of the 
effective transmission at the chemical potential. 
The results obtained using the  ``leading-log'' method only partly
agree to ours. While in our approach the one-parameter scaling
function depends on $K$ and the exponent $-2/K$ of the large
$x$-behavior is reproduced correctly to order $U$, the one-parameter
scaling function of the ``leading-log'' approach turns out to be
the non-interacting ($K=1$) function independently of the interaction
strength chosen.\cite{MatveevGlazman} This method thus does not capture 
all leading-order interaction effects.

This shortcoming of the ``leading-log'' method can be traced back to
the analytical form of the effective transmission at scale $\Lambda$ 
\begin{eqnarray}
\label{Glazmanform}
\left|t^{\Lambda}\right|^2= 
\frac{|t_0|^2 |\Lambda/\lambda_0|^{2 \alpha_B^l}}{|r_0|^{2 } +
|t_0|^2 |\Lambda/\lambda_0|^{2 \alpha_B^l} }
\end{eqnarray}
as it follows from the RG equations.\cite{MatveevGlazman}
The impurity parameters (height
and width) enter via the non-interacting transmission and reflection 
probabilities at the chemical potential
$|t_0|^2$ and $|r_0|^2$, and $\lambda_0$ denotes a non-universal energy scale. 
As above the exponent $\alpha_B^l$ stands for the leading-order term of
$\alpha_B=1/K-1$. 
This equation is derived under the
assumption that all energy scales, i.e.\ $T$, $\delta_N$, and the
energy of the incoming particle $\varepsilon$ (measured
relative to the chemical potential) are set to zero. 
The dependence on $T$, $\delta_N$, or $\varepsilon$ is then
determined by replacing $\Lambda$ by one of these variables.
For $T=0$, at the chemical potential, but for a finite system 
this gives
\begin{eqnarray}
\label{GlazmanformN}
|t(\delta_N)|^2= \frac{|t_0|^2 (\delta_N/\delta_0)^{2 \alpha_B^l}}{|r_0|^2 +
|t_0|^2 (\delta_N/\delta_0)^{2 \alpha_B^l} } \; , 
\end{eqnarray}
with the non-universal scale $\delta_0$.
Applying Eq.\ (\ref{LB}), which for $T=0$ reduces to $G=e^2
|t(\mu)|^2/h$, and introducing $x=|r_0/t_0| 
(\delta_N/\delta_0)^{- \alpha_B^l}$, where 
$ -\alpha_B^l$ is the leading-order term of $K-1$, the 
scaling function of the conductance turns out to be the 
non-interacting one, $\tilde G_{K=1} =1/(1+x^2)$. 
Eq.\ (\ref{GlazmanformN}) can be compared to our Eq.\ (\ref{ND1}) 
which applies to the situation of a single site impurity ($V_g=V$). 
Because of the 
above-mentioned equivalence of the $T$ and $\delta_N$ scaling it is
sufficient to consider $T=0$ and the length of the interacting wire 
as the variable. Without loss of generality we here focus on  
$n=1/2$ with $\mu =0$. For $\tilde t_l = \tilde t_r$, $T=0$, and 
at the chemical potential, i.e.\ for  
$\varepsilon = 0$, Eq.\ (\ref{ND1})  reads
\begin{equation} 
  |t(0,0,\delta_N)|^2 = \frac{4 \Delta^2(0,0,\delta_N)}{\left[ 
\tilde V + 2 \Omega(0,0,\delta_N)\right]^2 + 4 \Delta^2(0,0,\delta_N) } \; .
\label{ND1spec}
\end{equation} 
Because of the left-right symmetry we here and in the following
suppress the index
$l/r$. During the flow the potential $V$ on the impurity site
gets renormalized. At $\Lambda=0$ it is denoted by $\tilde V$.
Eq.\ (\ref{ND1spec}) holds for a more general impurity in which also 
the hopping amplitude $t_l=t_r$ [see the discussion following 
Eq.\ (\ref{hoimp})] to the left and right of the lattice site with 
onsite energy $V$ is different from one.\cite{footnotedotimp} 
$\Delta$ and $\Omega$ can be obtained numerically from the 
self-energy at $\Lambda=0$ as explained in Sect.\ \ref{Scattheory}. 
For $V \to 0$, which implies $\tilde V \to 0$ because of the 
particle-hole-symmetry for $V=0$, and sufficiently  large $N$
we find 
\begin{equation} 
\tilde V + 2 \Omega(0,0,\delta_N) \propto \delta_N^{\beta_w/2}\; , \;\;\;
\Delta(0,0,\delta_N) = \mbox{const.}  
\label{scalweak}
\end{equation}  
while for $V \gg B$ 
\begin{equation} 
\tilde V + 2 \Omega(0,0,\delta_N) = \mbox{const.}   \; , \;\;\;
\Delta(0,0,\delta_N) = \delta_N^{\beta_s/2} \; .  
\label{scalstrong} 
\end{equation}  
Instead of the one exponent $\alpha_B^l$ of the ``leading-log'' approach
we obtain two different exponents characterizing the weak- and
strong-impurity limits. This leads to an interaction-dependent scaling
function with a large $x$ scaling exponent which is correct to leading
order in $U$. In contrast to Eq.\ (\ref{GlazmanformN}) the analytical 
form Eq.\ (\ref{ND1spec}) allows for two different sources  of an 
$N$-dependence  of the effective transmission and the
conductance. We thus conclude that reducing the number of flowing
coupling constants to a single one --- the effective transmission at 
the chemical potential --- leads to an analytic form of the 
conductance which is too restricted to 
cover all important leading-order interaction effects. This shows 
that our fRG approach goes beyond the ``leading-log'' 
method.

\begin{figure}[htb]
\begin{center}
\includegraphics[width=0.3\textwidth,clip]{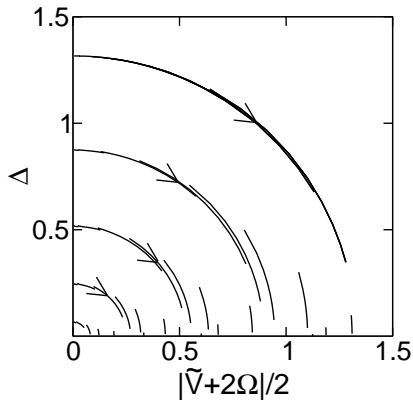}
\end{center}
\vspace{-0.3cm}
\caption[]{Flow diagram of the conductance in the 
plane defined by the real and imaginary part of the auxiliary Green
function.  The direction of the flow for $\delta_N \to 0$ and $U>0$ 
is indicated by the arrows. For $U<0$ it is reversed. The axis are
lines of fixed points. $\Delta$-axis: line of ``perfect chain'' fixed 
points with $G=e^2/h$.  $(\tilde V + 2 \Omega)$-axis: line of 
``open chain'' fixed points with $G=0$. 
\label{single7}}
\end{figure}

Eq.\ (\ref{ND1spec}) can be used to further visualize the scaling
of the conductance. Within our approximation the conductance can be
expressed in terms of the two scale- and parameter-dependent functions
$(\tilde V + 2 \Omega)/2$ and $\Delta \geq 0$.  Fig.~\ref{single7} shows 
a parameter plot in the $([\tilde V + 2 \Omega]/2, 
\Delta)$ plane
obtained at $\varepsilon=T=0$ and for varying $N$.
Each curve was calculated for a fixed $V$ and $t_l=t_r$ at $U=1$
and half filling. 
A similar plot can be obtained for a large fixed $N$
calculating $(\tilde V + 2 \Omega)/2$ and $\Delta$ for different
$\Lambda >0$ considering $\Sigma^{\Lambda}$ as an effective scattering 
potential. With increasing $N$, i.e.\ decreasing energy scale,
the ``flow'' approximately follows a section of a circle centered
around the origin, with a radius which depends on the initial impurity 
parameters. For $U>0$ the ``flow'' is clock-wise.
For attractive interaction the direction is 
reversed. The axis with $(\tilde V + 2 \Omega)/2=0$ can be considered 
as a line of ``perfect-chain'' fixed points with $|t|^2=1$, 
while the axis with $\Delta=0$ represents
the line of ``open-chain'' fixed points with $|t|^2=0$. 
In accordance with  Eqs.~(\ref{scalweak}) and (\ref{scalstrong}),
close to the two lines of fixed points the ``flow'' is perpendicular
to the axis. 

\begin{figure}[htb]
\begin{center}
\includegraphics[width=0.4\textwidth,clip]{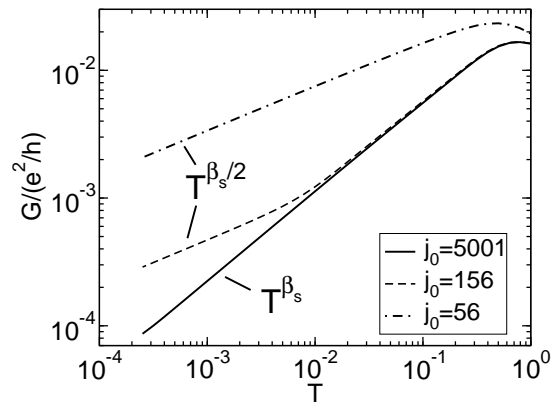}
\end{center}
\vspace{-0.3cm}
\caption[]{The conductance $G(T)$ for a site impurity with 
  $V=10$, $U=1$, $n=1/2$, $N=10^4+1$, and different impurity positions 
  $j_0$.\label{single8}}
\end{figure}

The above results are generic as long as the impurity is placed
sufficiently away from the contact regions. A richer behavior is 
found if the impurity is
positioned closer to the smooth contacts. Because of symmetry we only have 
to consider the case in which the impurity is moved towards the 
left contact. Then the scale
\begin{eqnarray}
\label{newscale}
\delta_{j_0} = \frac{\pi v_F}{j_0} 
\end{eqnarray} 
becomes important.\cite{FuNa} It sets the lower bound for the
power-law scaling with the exponents discussed above. For $T \approx
\delta_{j_0}$ we find a crossover to a power-law scaling with 
different exponents. This is exemplified in Fig.\ \ref{single8} 
for a strong impurity with $V=10$,
$N=10^4+1$, $U=1$ and three impurity positions 
$j_0=5001$ (solid line), $156$ (dashed line), $56$
(dashed-dotted line). For $j_0=156$ at $T 
\approx \delta_{j_0}$ the scaling with exponent $\beta_s$ 
crosses over to a scaling with exponent $\beta_s/2$. This coincides
with the prediction obtained  
within a field theoretical model where the exponent $2 \alpha_B$
crosses over to half its value $\alpha_B$.\cite{FuNa} For impurity 
positions in the contact region, i.e.~$j_0 = 56$, we only observe the 
exponent $\beta_s/2$ that describes the tunneling between 
the non-interacting leads and the TLL. 
The curve for $j_0 = 56$ does not follow the other two curves even at 
higher temperatures as compared to the bulk value the interaction at 
$j_0$ is reduced by a factor of two. 

Similar behavior is found for 
the scaling of  $1-G(T)/(e^2/h)$ in the limit of a weak impurity. 
In this case the 
exponent predicted for $\delta_N < T <\delta_{j_0}$ is
$2(K-1)/(K+1)$ and is thus different from half the exponent $2(K-1)$
found for $\delta_{j_0} < T \ll B$.\cite{FuNa} The exponent we find 
agrees to leading order in $U$ with this prediction. We note that 
even though $\beta_w/2$ agrees with $2(K-1)/(K+1)$ to order $U$ our
exponent in this regime is different from $\beta_w/2$.

\subsection{Resonant tunneling}
\label{resotun}

In this section we only investigate repulsive interactions $U > 0$ and
focus on $n=1/2$, with $\mu=0$, as the results for other fillings 
do not differ qualitatively. We furthermore only consider situation in
which the quantum dot is placed sufficiently away from the contact
regions. 

\subsubsection{$\bm{T=0}$}

\begin{figure}[htb]
\begin{center}
\includegraphics[width=0.4\textwidth,clip]{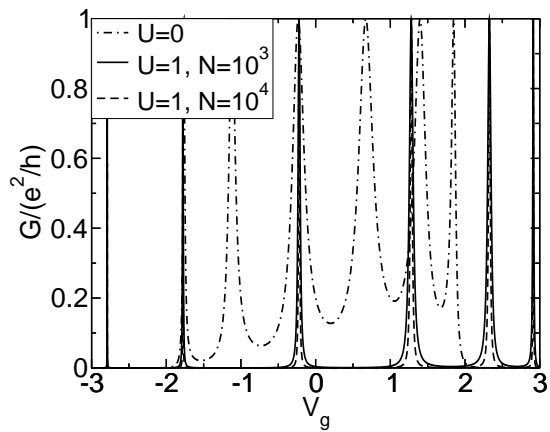}
\end{center}
\vspace{-0.3cm}
\caption[]{Gate-voltage dependence of the conductance for a symmetric 
double barrier at $T=0$. The parameters are $N_D=6$ and $V_{l/r}=2$. 
\label{double1}}
\end{figure}

We start our investigation of resonant tunneling considering $T=0$.
Fig.~\ref{double1} shows typical results for the gate-voltage $V_g$ 
dependence of the conductance for symmetric barriers. The quantum dot 
contains $N_D=6$ sites and is separated from the rest of the interacting 
wire by two site 
impurities of intermediate height $V_{l/r}=2$. 
In the figure the $N$-independent $U=0$ result (dashed-dotted line) is 
compared to the conductance of fRG calculations with $U=1$ for two different 
$N$. For not-too-weak barriers the number of perfect resonances 
(with peak conductance $e^2/h$) at positions $V_g^r$ coincides 
with the number $N_D$ of lattice sites which form the dot. For 
intermediate barriers and $U=0$ the peaks still overlap. For increasing 
$V_{l/r}$ the width of each peak decreases until well-separated 
resonances are formed. Also for $U>0$, $G$ 
shows perfect resonances with peak conductance $G_p=e^2/h$. 
This behavior can be understood 
from Eq.\ (\ref{NDres}), which for $U>0$ and sufficiently large 
$N$ can be used even for weak to intermediate barriers as 
the resonance
peaks do no longer
overlap. This was the necessary condition for the derivation of  Eq.\
(\ref{NDres}). 
At $V_g^{r}$ we find $\varepsilon_{\alpha} + 
2 \Omega^{(\alpha)}(0,0,\delta_N)=0$. Regardless of 
the $N$-dependence of $\Delta^{(\alpha)}(0,0,\delta_N)$ this leads
  to $|t(0,0,\delta_N)|^2=1$. At $T=0$ only $\varepsilon=0$ contributes in
  the Landauer-B\"uttiker formula (\ref{LB}), leading to
  $G=e^2/h$. Remarkably, $\Delta^{(\alpha)}(0,0,\delta_N)$ at $V_g^r$
does not follow the scaling $\delta_N^{\beta_s/2}$ of the spectral weight
close to a single impurity\cite{KaneFisher} but instead saturates
at small $\delta_N$. 
The imaginary part of the exact Green function,
which can be calculated from 
the auxiliary Green function using Eq.\ (\ref{Greenrel}), behaves
similarly. 
Increasing $U$ at fixed $N$ reduces the overlap between
the resonances by
increasing the energy difference between the peaks but even more 
importantly by reducing the width of each peak. The half-width $\sigma$ 
depends on  $N$. We find that the width of all peaks scales 
to zero as 
\begin{eqnarray}
\label{widthscalT_0}
\sigma \propto \delta_N^{-\beta_w/2} 
\end{eqnarray}
with the weak-single-impurity exponent $\beta_w$ as discussed in the 
last section. For $N \to \infty$ the resonances are thus infinitely sharp. 
It is important to note that for all parameter sets 
tested we found agreement between twice the scaling exponent of $\sigma$ 
at $T=0$ and the weak-single-impurity exponent 
up to our numerical accuracy. The difference of $\sigma$ for the
different resonance peaks at a fixed 
$N$ (see Fig.~\ref{double1}) is a band effect. Off resonance,
$G$ vanishes as 
\begin{eqnarray}
\label{offscalT_0}
G \propto \delta_N^{\beta_s} \; ,
\end{eqnarray} 
with the strong-single-impurity exponent $\beta_s$ (again up to the numerical 
accuracy). 
Using Eq.\ (\ref{NDres}) this behavior can be traced back to a 
$\Delta^{(\alpha)}(0,0,\delta_N) \propto \delta_N^{\beta_s/2}$
scaling while $\varepsilon_{\alpha} + 
2 \Omega^{(\alpha)}(0,0,\delta_N)$ goes to a constant 
different from 0.
Off resonance the double barrier acts as a single impurity of strength
$\Delta V_g=V_g-V_g^r$.\cite{KaneFisher} 
With this insight and using the one-parameter scaling ansatz Eq.\
(\ref{oneparascaling}) the $N$-dependence of $\sigma$, 
Eq.\ (\ref{widthscalT_0}) 
can be explained as follows: For small $|\Delta V_g|$ 
the non-universal scale (here $\delta_0$) goes as 
$|\Delta V_g|^{1/(1-K)}$. With $\tilde G_K(x_{1/2})=1/2$ 
this leads to $x_{1/2} =
c \delta_N^{K-1} \, |\Delta V_g^{(1/2)}|$, with a 
$\Delta V_g$-independent constant $c$. We thus find  
$\sigma = 2 |\Delta V_g^{(1/2)}|  \propto \delta_N^{1-K}$ which agrees with 
Eq.\ (\ref{widthscalT_0}) to leading order in $U$.

\begin{figure}[htb]
\begin{center}
\includegraphics[width=0.4\textwidth,clip]{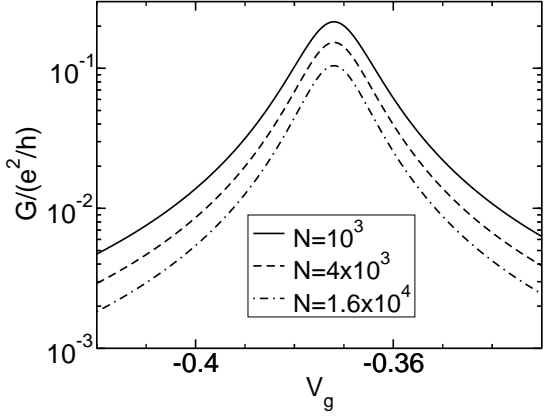}
\end{center}
\vspace{-0.3cm}
\caption[]{Gate voltage dependence of the conductance for an asymmetric 
double barrier at $T=0$. The parameters are $U=0.5$, $N_D=10$, 
$t_l=0.1$, and $t_r=0.3$. Only gate voltages close 
to $V_g^r$ are shown. Note the linear-log scale.\label{double2}}
\end{figure}

For asymmetric barriers and $U=0$ the conductance still displays
resonances but with peak values $G_p$ which are smaller than $e^2/h$. 
The same holds for $U>0$ as is shown in Fig.~\ref{double2} for
hopping barriers with $t_l=0.1$, $t_r=0.3$,
$U=0.5$, and $N_D=10$. 
We set the interaction across the barrier bonds to 0. The number of resonance 
peaks is $N_D$. Here we only show a single peak. 
For hopping impurities as barriers, $G$ is symmetric around $V_g=0$
(for $n=1/2$).
With increasing $U$ and for 
fixed $N$, $G_p$ decreases. The same holds for
fixed $U$ and increasing $N$. For sufficiently large $N$ we find that 
Eq.\ (\ref{offscalT_0}) holds for the off- and on-resonance scaling, 
i.e.\ on small scales also at $V_g^r$ an asymmetric double barrier acts 
as a single impurity. As a consequence for $N \to \infty$ the
half-width  $\sigma$
of the peaks approaches a constant.
Again Eq.\ (\ref{NDres}) can be used to further analyze $G$. 
At $V_g^{r}$, $\varepsilon_{\alpha} + 
\Omega_l^{(\alpha)}(0,0,\delta_N)
+ \Omega_r^{(\alpha)}(0,0,\delta_N) =0$.
For $t_r <t_l <1$ and $\delta_N \to 0$, $\Delta_{r}(0,0,\delta_N)$ vanishes as 
$\delta_N^{\beta_s/2}$ while $\Delta_{l}(0,0,\delta_N)$ increases as 
$\delta_N^{-\beta_s/2}$. These results combine to Eq.\ (\ref{offscalT_0}).
Applying Eq.\ (\ref{Greenrel}) one can show that in contrast to the
asymmetric behavior of the imaginary part $\Delta_{l/r}$ of the
auxiliary Green function, the spectral
function of the exact Green function on the sites to the right and the
left of the barriers both scale to zero as $\delta_N^{\beta_s/2}$.   
The off-resonance behavior can be explained 
similarly as for symmetric barriers: $\Delta_{l/r}(0,0,\delta_N) \propto 
 \delta_N^{\beta_s/2}$ while $\varepsilon_{\alpha} + 
\Omega_l^{(\alpha)}(0,0,\delta_N)
+ \Omega_r^{(\alpha)}(0,0,\delta_N)$ goes to a constant different from 0. 
We obtain all the above results for both hopping and site impurities
as barriers.

Within an extension of the LSGM to the two-barrier case resonant
tunneling on asymptotically small scales has been discussed in Ref.\
\onlinecite{KaneFisher}. The results of that work agree with ours. 
As for the single-impurity case we reproduce the exponents
of the power-laws (\ref{widthscalT_0}) and (\ref{offscalT_0}) to
leading order in $U$, i.e.\ the exponents of the extended 
LSGM are given by $1-K$ and $2 \alpha_B$, respectively. 
In the low-energy limit, the conductance exhibits the same scaling
behavior as a function of $T$ or $\delta_N$, but for higher
temperatures
the $T$ dependence of $G$ at fixed $N$ is much richer
than the $\delta_N$ scaling, showing 
non-monotonic behavior and power-laws with different 
exponents in a variety of temperature regimes. 

\subsubsection{$\bm{T>0}$}

Shortly after the discussion of resonant tunneling on asymptotically
small scales the conductance for intermediate to high
energies was investigated within a field-theoretical model 
using different approximate
methods.\cite{Furusaki1} For symmetric barriers and applying
second-order perturbation theory in the barrier height the $T>0$ deviations 
of $G_p(T)$ (at resonance) from $e^2/h$ were determined to scale as 
$T^{2K}$.\cite{Furusaki1} At further increasing $T$ a regime of
uncorrelated sequential tunneling (UST) was predicted 
based on a perturbative analysis in the inverse barrier height, 
where $G_{ p}(T) \propto T^{\alpha_{ B}-1}$ and $\sigma(T) \propto
T$.\cite{Furusaki1,Furusaki2} This regime is bounded from above by 
the (non-interacting) level spacing of the dot $\delta_{N_D}=\pi v_{ F}/N_D$.
Within the same perturbative approach and for 
$\delta_{N_D} \ll T \ll B$, $G_p(T)$ increases as $T^{2 \alpha_B}$ 
for increasing $T$.\cite{Furusaki1}  
In contradiction to the temperature 
dependence following from UST, in recent transport experiments on 
carbon nanotubes a  suppression of  $G_p$ with decreasing $T$ was 
observed for $T \lessapprox \delta_{N_D}$.\cite{Postma}  
Using another approximation scheme,  
``correlated sequential tunneling'' (CST) was predicted to replace 
UST.\cite{Thorwart,Thorwart2} It leads to $G_{ p} \propto T^{2
  \alpha_{ B}-1}$. Although a simplified field-theoretical model of
spinless fermions was used and contacts as well as 
leads were ignored the CST scaling was argued to be consistent
with the experimental data. This consideration has stimulated a number of
theoretical works.\cite{Komnik,Nazarov,Polyakov,Huegle} 
In particular, Quantum Monte Carlo (QMC) results\cite{Huegle} for 
$G_p(T)$ were interpreted to be consistent with CST for 
$T \lessapprox \delta_{N_D}$. Resonant tunneling was also investigated
using the ``leading-log'' method.\cite{Nazarov,Polyakov} Within this
approach no indications of a CST regime were found.

\begin{figure}[htb]
\begin{center}
\includegraphics[width=0.47\textwidth,clip]{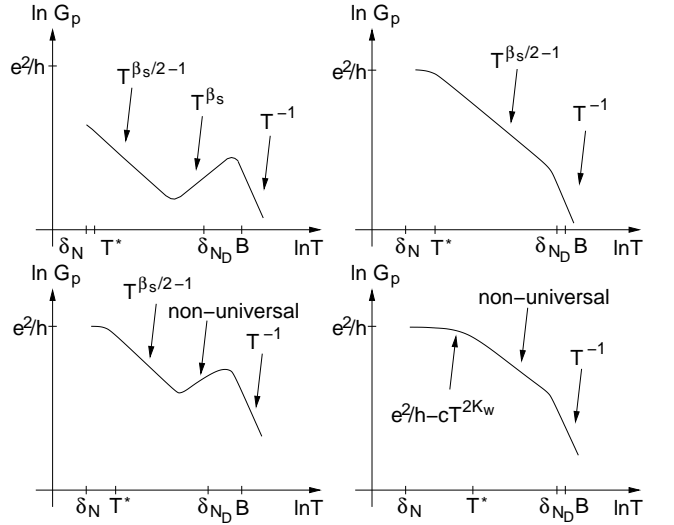}
\end{center}
\vspace{-0.3cm}
\caption[]{Schematic plot of the different regimes for the scaling of 
  the conductance $G_p$ at resonance for symmetric barriers found
  using our approximation scheme. The upper left panel is for dots
  with a large number of lattice sites $N_D$ and high barriers, 
  the upper right for small $N_D$ and high barriers, the lower left 
  for large $N_D$ and low barriers, and the lower right for small 
  $N_D$ and low barriers. The relevant energy scales are the band
  width $B$, the $U=0$ level spacing of the dot $\delta_{N_D}$, the
  lower bound of the UST regime $T^\ast$ [see Eq.~(\ref{Tastdef})], and the
  infrared cutoff $\delta_N$ associated with the number of lattice 
  sites forming the interacting wire. 
\label{skizzereso}}
\end{figure}

Being applicable for any barrier height, but only for interactions 
such that $1/2 \leq K \leq 1$, our fRG based method is complementary
to the approaches of Refs.\ \onlinecite{Furusaki1} and 
\onlinecite{Furusaki2}. It has the advantage that all relevant 
temperature regimes can be investigated within one approximation 
scheme (see Ref. \onlinecite{VM6} and below). 

In Fig.~\ref{skizzereso} our findings for 
$G_p(T)$ and a fixed $N$ presented in Ref.\ \onlinecite{VM6} are summarized 
schematically in the four limiting cases of small and large quantum
dots separated from the rest of the interacting wire by high and low
barriers. The $T^{-1}$ behavior found for scales of the order of $B$
and larger is the same band effect as discussed above for the
single-impurity case. 

For large dots and high barriers (upper left
panel of Fig.~\ref{skizzereso}) the $T^{-1}$ behavior
is followed by a $T^{\beta_s}$ scaling, with the strong-single-impurity
exponent $\beta_s$, down to temperatures slightly less than
$\delta_{N_D}$. According to Eq.\ (\ref{Kirchhoff}) in this parameter and 
temperature regime $G_p(T)$ is obtained by adding the resistances 
of the two individual barriers (Kirchhoff's law). Both follow the 
single impurity 
scaling  $T^{\beta_s}$  which explains that the same exponent is found
in resonant tunneling. For even lower $T$, $G_p$ follows a power-law 
with (to leading order in $U$) the UST exponent $\beta_s/2-1$ down
to a scale $T^*$. 
The crossover from the regime in which Kirchhoff's law can be applied
to the UST regime is relatively sharp and takes typically half an order
of magnitude (see Fig.~2 of Ref.\ \onlinecite{VM6}).  
For $T \ll \delta_{N_D}$ the width of 
$-df/d\varepsilon$ is smaller than $\delta_{N_D}$ and only a single
resonance peak (the one around 0) of 
$|t(\varepsilon,T,\delta_N)|^2$ contributes to the integral in 
Eq.\ (\ref{LB}). The width $\Delta^{(\alpha)}$  
of this peak  
vanishes as $\tau^2 T^{\beta_s/2}/N_D$ leading to $G_p(T) \propto
T^{\beta_s/2-1}$ as discussed in Sect.~\ref{Scattheory}. 
In the case of hopping barriers $\tau $ stands for $t_{l/r}$ 
while it is proportional to $1/V_{l/r}$ for site barriers.
The lower 
bound of this scaling regime is reached when $T$ is equal to 
$\Delta^{(\alpha)}$, i.e.\ at 
\begin{eqnarray}
\label{Tastdef}
T^\ast \propto
(\tau^2/N_D)^{1/(1-\beta_s/2)} \; .
\end{eqnarray}
Using Eq.\ (\ref{LB}) it is easy to see that for 
$T^\ast \ll T \ll \delta_{N_D}$ the half-width $\sigma$ of 
the resonance in $G(T)$ as a function of $V_g$ scales like $T$.
These results hold for all cases in which we find a UST regime,
i.e.\ for the other dot parameters discussed next, but also for
asymmetric barriers and the off-resonance scaling investigated further
below.  For large dots, high barriers, and the values 
of $N$ used here, $T^{\ast}$ is very close
to the scale $\delta_N$. Below this scale no interaction-dependent
``universal'' behavior can be observed.   

\begin{figure}[htb]
\begin{center}
\includegraphics[width=0.4\textwidth,clip]{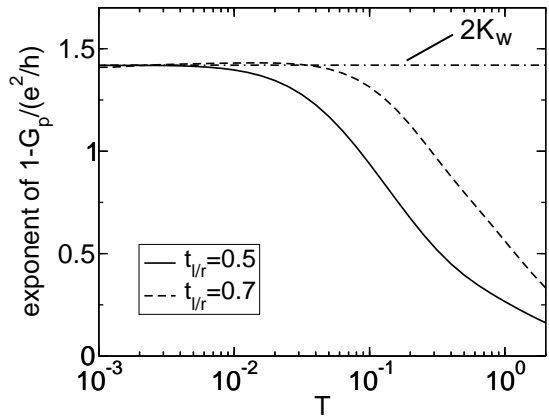}
\end{center}
\vspace{-0.3cm}
\caption[]{Effective exponent (i.e.\ logarithmic derivative) 
of $1- G_p(T) /(e^2/h)$ for a small dot
  with $N_D=1$, $U=1$, $N=10^4$, and intermediate to weak
  hopping barriers $t_{l/r}$. Dashed-dotted line: $2K_w$.
\label{double2a}}
\end{figure}

For small dots and high barriers (upper right panel of Fig.\
\ref{skizzereso}) $\delta_{N_D}$ and $B$ are very close to
each other and no regime with $T^{\beta_s}$ scaling is developed.
Due to the factor $1/N_D$ in Eq.\ (\ref{Tastdef}), for small
dots $T^\ast$ is larger than for large $N_D$.
For small dots a UST regime with exponent $\beta_s/2-1$ is not as
nicely developed as for large $N_D$ (see Fig.~2 of Ref.~\onlinecite{VM6}
and Figs.\ \ref{double3}, \ref{double4}, and \ref{double5} below).

For large dots but low barriers (lower left panel of Fig.\
\ref{skizzereso}) $T^\ast$ moves to higher energies because of the 
factor $\tau$ in Eq.\ (\ref{Tastdef}). Between $B$ and $\delta_{N_D}$, $G_p$
decreases with decreasing temperature. As is discussed in Ref.\
\onlinecite{VM6} this decrease is not described
by a power-law. We thus denote it as non-universal behavior. 

For small $N_D$ and low barriers (lower right panel of Fig.\
\ref{skizzereso}) $\delta_N$ and $T^\ast$ are
sufficiently separated so that we can observe an additional
temperature regime with power-law scaling of $G_p$ and an
interaction-dependent exponent as exemplified in Fig.~\ref{double2a}
for $U=1$, $N=10^4$, $N_D=1$, and the barrier heights $t_{l/r}
=0.5$ as well as $t_{l/r} =0.7$. We find that for $\delta_N \ll T \ll T^\ast$, 
$G_p(T)$ approaches $e^2/h$ as
\begin{eqnarray}
\label{smallT}
1- G_p(T) /(e^2/h)  \propto T^{2 K_w}\; ,
\end{eqnarray}
with the fRG approximation $K_w$ for $K$ obtained from $\beta_w$ (see
Sect.\ \ref{Singimp}). In this temperature regime
and for energies $\varepsilon$ which contribute to the 
integral in Eq.\ (\ref{LB}), $|t(\varepsilon,T,\delta_N)|^2$ does 
not depend on $T$. We find that for $\delta_N \ll \varepsilon
\ll T^\ast$, $\varepsilon  -\varepsilon_{\alpha} 
- 2 \Omega^{(\alpha)}(\varepsilon,0,\delta_N) \propto 
\varepsilon$ and $\Delta^{(\alpha)}(\varepsilon,0,\delta_N) \propto
\varepsilon^{-\beta_w/2}$. Evaluating the $\varepsilon$ integral for
$\delta_N \ll T \ll T^\ast$ leads to Eq.\ (\ref{smallT}). 
For low barriers, small $N_D$, and $T^\ast \ll
T \ll \delta_{N_D}$ we no longer find the UST power-law but instead 
non-universal behavior.

For the cases in which a comparison is possible the above exponents agree
(to leading order in $U$) with the results obtained for a
field-theoretical model using perturbation theory in the barrier
height and the inverse barrier height.\cite{Furusaki1,Furusaki2} In addition 
applying the fRG to our microscopic model we can describe the
complete crossover between the temperature regimes with ``universal''
power-laws and cover the parameter regimes with  non-universal   
behavior. 

We do not find any indications of the CST exponent 
$2 \alpha_B-1$. Besides the single-and double-barrier problems some of
us also studied transport of correlated electrons through junctions of
$M$ 1d wires, e.g.\ Y-junctions with $M=3$ and X-junctions with
$M=4$.\cite{Xavier} Also in these geometries one finds power-law 
scaling of the conductance. For a certain case (Y-junction with 
magnetic flux) the $K$-dependent exponent is known 
analytically,\cite{Claudio} which we again reproduce to leading 
order in $U$.\cite{Xavier}  
We thus conclude that in all cases of inhomogeneous TLLs we have
studied and in which exact results for exponents (from effective
field-theoretical models) of power-law scaling are known we
reproduce these exponents at least to leading order. 
We find it thus very likely that the occurrence of the CST regime 
with exponent $2 \alpha_{B}-1$ is an artifact of the approximations 
used in Refs.\ \onlinecite{Thorwart} and \onlinecite{Thorwart2}.

As Fig.~3 of Ref.\ \onlinecite{VM6} shows, the non-universal behavior
discussed above might quite easily be identified incorrectly as 
power-law scaling with an exponent, which is significantly 
smaller than $\beta_s \approx 2 \alpha_B$, i.e.\ in the vicinity 
of $2 \alpha_B -1$. This is of 
particular importance for the interpretation of noisy QMC data as 
well as noisy experimental results, both typically restricted to a
temperature regime of one to two orders of magnitude.\cite{Huegle,Postma} 
In Fig.~4 of Ref.\ \onlinecite{Huegle} a single set of QMC data 
with $K=0.6$, for an intermediate dot size, 
and intermediate barriers is presented roughly falling into the
parameter regime shown in the lower left panel of Fig.\
\ref{skizzereso}.  In  Ref.\
\onlinecite{Huegle} a certain part of the
temperature regime with decreasing $G_p(T)$ (for decreasing $T$) was
fitted by a power-law with exponent $2 \alpha_B-1$ which was then 
interpreted to support the occurrence of CST. 

\begin{figure}[htb]
\begin{center}
\includegraphics[width=0.47\textwidth,clip]{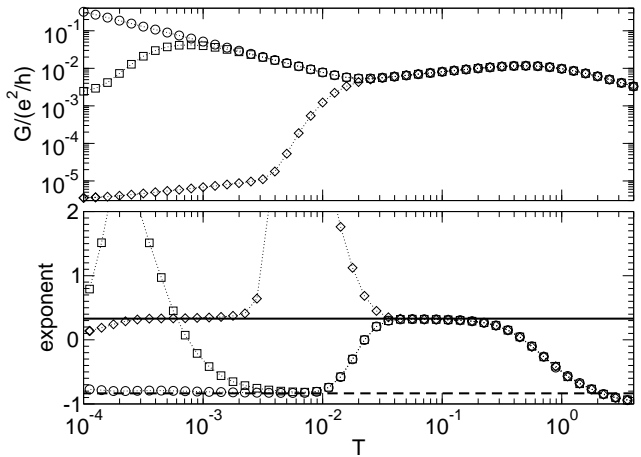}
\end{center}
\vspace{-0.3cm}
\caption[]{Upper panel: $G(T)$ for symmetric barriers with $V_{l/r}=10$ 
for $U=0.5$, $N=10^4$, $N_D=100$ at resonance $|\Delta V_g| =0$ (circle),
very close to a resonance with $|\Delta V_g| = 0.001$ (squares),
and  in a conductance minimum with $|\Delta V_g| =  0.04$ (diamonds).
Lower panel: Logarithmic derivative of $G(T)$. Solid line: 
$\beta_s$; dashed line: $\beta_s/2 -1$.  \label{double3}}
\end{figure}

\begin{figure}[htb]
\begin{center}
\includegraphics[width=0.47\textwidth,clip]{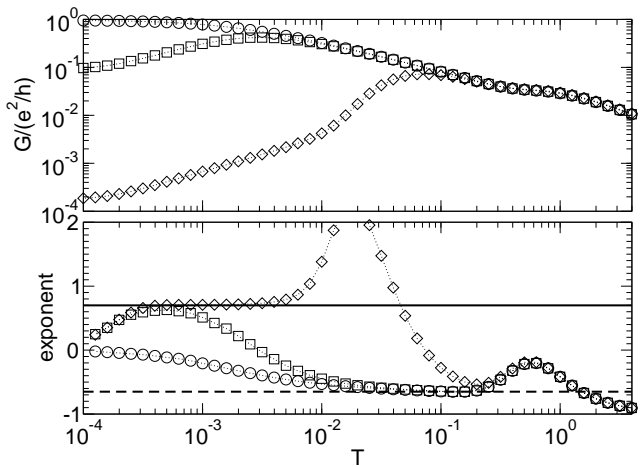}
\end{center}
\vspace{-0.3cm}
\caption[]{Upper panel: $G(T)$ for symmetric barriers with $t_{l/r}=0.2$ 
for $U=1$, $N=10^4$, $N_D=4$ at resonance $|\Delta V_g| =0$ (circle),
close to a resonance with $|\Delta V_g| = 0.005 $ (squares),
and further away from the resonance with $|\Delta V_g| = 0.1 $ (diamonds).
Lower panel: Logarithmic derivative of $G(T)$. Solid line: 
$\beta_s$; dashed line: $\beta_s/2 -1$.  \label{double4}}
\end{figure}

We next discuss the $T$ dependence of the conductance away from
$V_g^r$ still considering symmetric barriers. In the upper panel 
of Fig.~\ref{double3} $G(T)$ is shown 
for $U=0.5$, $N=10^4$, a large dot with $N_D=100$, high barriers with 
$V_l=V_r=10$, and three different $|\Delta V_g|$. Results for the
resonance closest to $V_g=0$ are shown. We have verified that 
similar behavior can be found for the other resonances.
The circles show the conductance at  $|\Delta V_g|=0$. The peak
conductance $G_p(T)$ displays the power-laws discussed above. This is
best seen in the lower panel in which the logarithmic derivative of
the conductance is shown. Close to the resonance (squares) for 
decreasing $T$, 
$G(T)$ follows $G_p(T)$ into the UST regime before the conductance at 
$|\Delta V_g| \neq 0 $ decreases while $G_p(T)$ tends towards $e^2/h$. 
For $|\Delta V_g| \neq 0$ 
the peak in $|t(\varepsilon,T,\delta_N)|^2$ is centered around 
$\Delta V_g$. In the UST regime its width still scales as 
$\Delta^{(\alpha)} \propto \tau^2 T^{\beta_s/2}/N_D$  
(see Eq.\ (\ref{NDres}) and above).   
As long as $|\Delta V_g| \ll \delta_{N_D}$  
the curves separate when $\Delta^{(\alpha)}$ is equal
to $|\Delta V_g|$. This leads to the ``off-resonance'' temperature 
scale
\begin{eqnarray}
\label{sepscale}
T_{\rm or} \propto \left( N_D |\Delta V_g|/\tau^2 \right)^{2/\beta_s}
\; ,
\end{eqnarray}   
at which the off-resonance conductance separates from $G_p(T)$.
Sufficiently away from the resonance, e.g.\ in the conductance minimum 
(diamonds), the temperature regime in which Kirchhoff's law can be 
applied is followed by a regime in which the double barrier acts 
as a strong single impurity. This crossover typically takes one order 
of magnitude in $T$ and consists of a drop of $G$ by at least two orders 
of magnitude. 
In both these regimes $G(T)$ follows a power-law with exponent $\beta_s$, 
which arises for completely different reasons. The origin of the power-law 
for $\delta_{N_D} \leq T \ll B$ has been explained above. For these
temperatures many peaks in  $|t(\varepsilon,T,\delta_N)|^2$ contribute
to the $\varepsilon$ integral in Eq.\ (\ref{LB}). 
The single-impurity scaling on 
asymptotically small scales can be understood as follows: Off
resonance  and close to the chemical 
potential $|t(\varepsilon,T,\delta_N)|^2$ behaves like 
$\left(\varepsilon^2 + T^2\right)^{\beta_s/2}$ (for $\varepsilon,T \gg
\delta_N$). 
In the evaluation of the integral in Eq.\
(\ref{LB}) the derivative of the Fermi function can be considered as
being a constant (as a function of $\varepsilon$) which scales like
$1/T$. The remaining integral with integrand 
$\left(\varepsilon^2 + T^2\right)^{\beta_s/2}$ runs over
an energy range of order $2 T$ which then leads to $G(T) \propto 
T^{\beta_s}$.
For $T \lessapprox \delta_N/2$ a deviation from the 
described power-law scaling sets in.  
As an additional example for the off-resonance behavior, in Fig.\
\ref{double4} we show $G(T)$ for $U=1$, $N_D=4$, $N=10^4$, weak
hopping barriers with $t_{l/r}=0.2$ (again the interaction across the barriers
is set to 0) and three different $|\Delta V_g|$.

\begin{figure}[htb]
\begin{center}
\includegraphics[width=0.47\textwidth,clip]{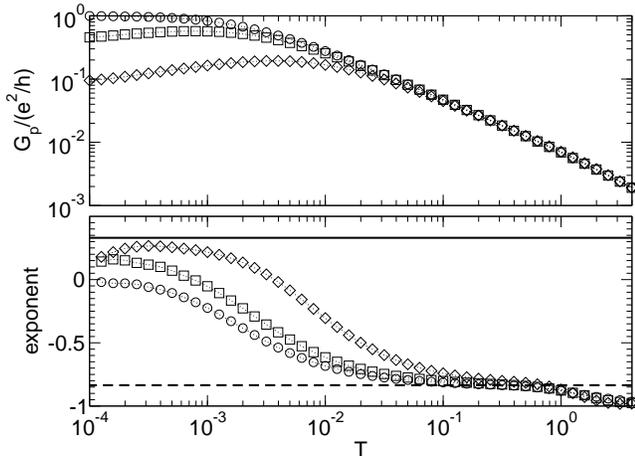}
\end{center}

\vspace{-0.7cm}

\caption[]{Upper panel: $G_p(T)$ for
asymmetric barriers with $U=0.5$, $N=10^4$, $N_D=2$, and $V_l=7$,
$V_r=12.3$ (squares) as well as $V_l=4$, $V_r=13.6$ (diamonds).
For comparison also the symmetric case with 
$V_l=V_r=10$ is shown (circles).
Lower panel: Logarithmic
  derivative of $G_p(T)$. Solid line: $\beta_s$; dashed
  line: $\beta_s/2 -1$.  \label{double5}}
\end{figure}

\begin{figure}[htb]
\begin{center}
\includegraphics[width=0.47\textwidth,clip]{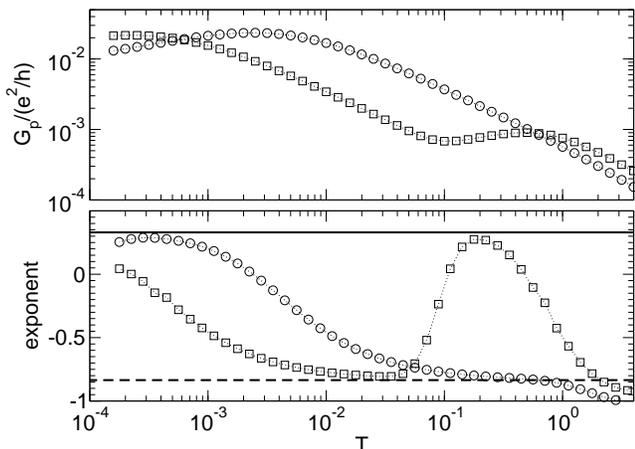}
\end{center}

\vspace{-0.7cm}

\caption[]{Upper panel: $G_p(T)$ for
asymmetric barriers with $V_{l}=5$, $V_r=50$, $U=0.5$, $N=10^4$, 
$N_D=2$ (circles) and $N_D=20$ (squares). Lower panel: Logarithmic
  derivative of $G_p(T)$. Solid line: $\beta_s$; dashed
  line: $\beta_s/2 -1$.  \label{double6}}
\end{figure}

We finally describe the $T$-dependence of the conductance 
for asymmetric barriers at resonance. In Fig.~\ref{double5} results 
are presented for $U=0.5$, $N_D=2$, $N=10^4$, and different site 
barriers with weak to intermediate asymmetry $V_l=7$, $V_r=12.3$
(squares) and $V_l=4$, $V_r=13.6$ (diamonds). The combination of
$V_{l/r}$ was chosen such that at high temperatures the curves
approximately collapse. For comparison also the symmetric case with 
$V_l=V_r=10$ is shown (circles). Over a wide range of temperatures 
$G_p(T)$ for asymmetric barriers shows the same behavior as in the
symmetric case. The scale $T_{\rm as}$ on which $G_p(T)$ for 
$V_l \neq V_r$ starts to deviate from the behavior of the symmetric 
case discussed above decreases with vanishing asymmetry parameter 
$\gamma=|V_l^{2} - V_r^{2}| /(V_r^{2} + V_l^{2})$. 
In addition $T_{\rm as}$ decreases for 
increasing $N_D$ as can be seen in Fig.~\ref{double6} which shows 
$G_p(T)$ for $U=0.5$, $V_l=5$, $V_r=50$, $N=10^4$ and two different
$N_D$. Even for a fairly strong asymmetry with $\gamma \approx 0.98 $ 
($V_l=5$, $V_r=50$) for $N=10^4$ we observe the asymptotic low-energy 
behavior $G_p(T) \propto T^{\beta_s}$ only for very small $N_D$ 
(see Fig.~\ref{double6}). Already for $N_D=20$, $T_{\rm as} < \delta_N$.  

The temperature dependence of the conductance for asymmetric barriers
and $V_g \neq V_g^r$ can be deduced from $G_p(T)$ obtained  in the
last paragraph (asymmetric barriers) and $G(T)$ for symmetric 
barriers at $|\Delta V_g| > 0$ discussed further above.

\section{Summary and perspectives}
\label{Summary}

In this paper we presented a method to calculate
the linear conductance through inhomogeneous TLLs connected to semi-infinite
non-interacting leads. The system is described by a microscopic
lattice model of spinless fermions. As applications we studied
transport in the presence of a single impurity as well as the
double-barrier problem, allowing for resonant tunneling. The contacts
are modeled to be smooth, i.e.\ in the absence of impurities and for
$T \ll B$ the conductance is equal to $e^2/h$. This requires that the
interaction is turned on smoothly close to the two contacts. 
The transport problem was previously investigated using effective
field-theoretical models. In the cases where exact results for such
models are known we obtained quantitative agreement for weak to intermediate
interactions with $1/2 \leq K \leq 1$. At half-filling this
corresponds to nearest-neighbor interactions $0 \leq U \leq 2$, which
covers the whole parameter regime in which the model is a TLL (with $K
<1$). Away from half-filling even larger $U$ can be considered. 

Our method captures the
expected power-law scaling of the conductance for a single weak and
strong impurity. In addition, we can describe the complete crossover
governed by a $K$-dependent one-parameter scaling function. 
For a single impurity we briefly studied attractive interactions. Also
in this case our method provides reliable results for not-too-strong 
interactions with $1 \leq K \leq 3/2$, i.e.~at half-filling for $-1
\leq U \leq 0$. 

For resonant tunneling depending on the parameters of the quantum dot
we find several temperature regimes with power-law scaling as well as
non-universal behavior. All these temperature regimes are obtained
within the same approximation scheme. The crossover between the
regimes can be studied in detail. For parameters for which 
a comparison is possible our results agree with the ones obtained in 
lowest order perturbation theory in the barrier height and inverse 
barrier height. We do not find any indications of a 
``correlated sequential tunneling'' regime with the exponent $2
\alpha_B-1$ predicted from an approximate Master-equation
approach.\cite{Thorwart,Thorwart2} 
If it would be present our method should be able to reveal such a 
regime with a scaling exponent which is of leading order in the 
interaction. 

Our method is a very flexible tool to investigate models of 
inhomogeneous TLLs. It allows for extensions which are 
required to obtain a more realistic description of experimental systems:
(i) The spin degree of freedom can be included. (ii) Models with more
realistic contacts (extended bulk and end contacts) can be treated. 
After projecting out the non-interacting leads the microscopic details 
of the contacts enter the fRG flow equations as the initial condition
of the self-energy. Preliminary results show that imperfect leads
have a strong effect on $G(T)$ even in cases in which the imperfection
is fairly weak. (iii) More realistic models for the leads can be
used. All that enters the fRG approach is the local Green function of
the lead at the contact after disconnecting the lead and interacting
wire. In addition the transport through junctions of interacting wires, such
as $Y$- and $X$-junctions, as well as transport through Aharonov-Bohm 
geometries can be described using the fRG.

\section*{Acknowledgments}

We thank M.~Salmhofer and 
H.~Schoeller for valuable discussions and U.~Schollw\"ock for
collaboration during the early stage of this research project. 
While completing this work all of us attended the {\it Winter School
  on Renormalization Group Methods for Interacting Electrons} at the {\it
  International Center of Condensed Matter Physics} of the University of
Bras\'ilia, Brazil. We benefited from the possibility to discuss the
issues presented here during the workshop. We thank 
A.~Ferraz and the staff of the ICCMP for their hospitality. 
X.B.-T. was supported by a 
Lichtenberg-Scholarship of the 
``G\"ottingen Graduate School of Physics''.

\section*{Appendix}
\label{vertexcor}

The Kubo formula for the conductance has a contribution with bare 
current operators and
one with current-vertex corrections.  In
this appendix we explain why the latter vanish in our approximation,
consistent with the Ward identity.

We have approximated the full two-particle interaction vertex $\Gamma$
by a renormalized nearest-neighbor interaction which is independent of
frequency.  Then the flow of the right
current vertex $\Lambda_R(\omega+i0)$ is of order $\omega$ and vanishes
in the limit of the dc conductance.  The approximation that $\Gamma$
is frequency independent has another consequence: by the flow
equation, it follows that $\mbox{Im} \, \Sigma = 0$, that is we do not
capture inelastic processes at second order in the interaction.  They
could be included in the flow equation by retaining the frequency
dependence of $\Gamma$.

However, the fact that the vertex corrections and $\mbox{Im} \,
\Sigma$ vanish simultaneously shows that our approximation is at least
consistent with the Ward identity associated with  particle
number (charge) conservation.  The global continuity equation for the
interacting region is $\partial n_C / \partial t + J_R - J_L = 0$,
where $n_C$ is the total particle number in the interacting region,
and $J_{L,R}$ are the currents entering the system from the left and
leaving it to the right.  This continuity equation implies a 
Ward identity.\cite{Oguri}  After analytical continuation to frequencies
slightly above and below the real axis (the branch cut of the current
operators and the self-energy), the density response drops out and we
obtain
\begin{eqnarray*}
   P_R(\epsilon+i0,\epsilon-i0) - P_L(\epsilon+i0,\epsilon-i0)
   = 2 \, \mbox{Im} \, \Sigma(\epsilon+i0) \,,
\end{eqnarray*}
where $P_{L,R}$ are the corrections of the left and right current
vertices, using the notation of Ref.~\onlinecite{Oguri}.  Thus, the
current-vertex corrections are related to the discontinuity across the
real axis of the
self-energy.  At $T=0$, both sides vanish exactly because there is no
inelastic scattering, while at finite temperatures, neglecting
inelastic processes is a consistent approximation.


\begin{thebibliography}{*}
\bibitem{footnoteleads} We consider spinless TLLs connected to semi-infinite
  Fermi-liquid leads, which implies that the impurity free conductance
  is given by $e^2/h$.
\bibitem{residue} This can easily be shown by calculating the
  necessary integral by the theorem of residues.
\bibitem{footnotet_lr} Due to renormalization 
the effective hopping matrix elements $\tilde t_{l/r}$ considered 
in Sect.\ \ref{Scattheory} are not necessarily the same as the bare hoppings 
$t_{l/r}$ of Sect.\ \ref{model}.
\bibitem{footnoteV_g} Due to renormalization 
the effective gate voltage $\tilde V_g$ 
considered in Sect.\ \ref{Scattheory} is
not necessarily the same as the bare gate voltage $V_g$ of 
Sect.\ \ref{model}.
\bibitem{footnotealgo} It is a finite $T$ generalization of the 
$T=0$ algorithm presented in Ref.\ \onlinecite{Sabine}.
\bibitem{footnotenonperfect} We will present a detailed study of the 
role of more realistic contacts in an upcoming publication.
\bibitem{footnotedotimp} For $\tilde V \neq 0$, i.e.\ off resonance, 
the resulting single site dot can be understood as a single impurity.
\end{thebibliography}
\end{document}